\documentclass[bibyear]{aa} 

\PassOptionsToPackage{colorlinks=true,allcolors=blue,breaklinks=true}{hyperref}
\usepackage[switch]{lineno}

\usepackage{subcaption}
\usepackage{graphicx}
\usepackage{multirow}

\usepackage[colorlinks=true, allcolors=blue]{hyperref}

\usepackage{natbib,twoopt}

\usepackage[breaklinks=true]{hyperref} 
\bibpunct{(}{)}{;}{a}{}{,}             
\makeatletter
  \newcommandtwoopt{\citeads}[3][][]{\href{http://adsabs.harvard.edu/abs/#3}%
    {\def\hyper@linkstart##1##2{}%
     \let\hyper@linkend\@empty\citealp[#1][#2]{#3}}}
  \newcommandtwoopt{\citepads}[3][][]{\href{http://adsabs.harvard.edu/abs/#3}%
    {\def\hyper@linkstart##1##2{}%
     \let\hyper@linkend\@empty\citep[#1][#2]{#3}}}
  \newcommandtwoopt{\citetads}[3][][]{\href{http://adsabs.harvard.edu/abs/#3}%
    {\def\hyper@linkstart##1##2{}%
     \let\hyper@linkend\@empty\citet[#1][#2]{#3}}}
  \newcommandtwoopt{\citeyearads}[3][][]%
    {\href{http://adsabs.harvard.edu/abs/#3}
    {\def\hyper@linkstart##1##2{}%
     \let\hyper@linkend\@empty\citeyear[#1][#2]{#3}}}
\makeatother

\let\oldthebibliography\thebibliography
\renewcommand{\thebibliography}[1]{%
  \oldthebibliography{#1}
  \let\oldbibitem\bibitem
  \let\oldtextsc\textsc
  \def\oldbbland{et}
  \newcounter{authorcount}
  \def\bibitem[##1]##2{%
    \let\textsc\oldtextsc
    \let\bbland\oldbbland
    \oldbibitem[##1]{##2}%
    \let\textsc\mytextsc%
    \let\bbland\mybbland
    \setcounter{authorcount}{0}
  }
  \def\mybbland{\setcounter{authorcount}{0}\oldbbland}
  \def\dropetal##1.{ \bbletal}
  \def\mytextsc##1{%
    \oldtextsc{##1}%
    \stepcounter{authorcount}%
    \ifnum\value{authorcount}=3\relax%
      \expandafter\dropetal%
    \fi%
  }%
}

\usepackage{txfonts}

\newcount\Comments  
\newcommand{\kibitz}[2]{\ifnum\Comments=1\textcolor{#1}{#2}\fi}

\begin{document} 

   \title{Optical spectral characterization of OP 313}

   \subtitle{Constraining the contribution of thermal and non-thermal optical emission}

   \author{J. Otero-Santos\fnmsep\thanks{\email{jorge.otero@pd.infn.it}}
          \inst{1}
          \and
          M. Nievas Rosillo\inst{2,3}
          \and
          J. A. Acosta-Pulido\inst{2,3}
          \and
          R. Clavero\inst{2,3}
          }

   \institute{Istituto Nazionale di Fisica Nucleare, Sezione di Padova, 35131 Padova, Italy
            \and
             Instituto de Astrof\'isica de Canarias (IAC), E-38200 La Laguna, Tenerife, Spain 
            \and
             Universidad de La Laguna (ULL), Departamento de Astrof\'isica, E-38206 La Laguna, Tenerife, Spain 
             }

   \date{Received December 23, 2025; accepted February 4, 2026}
 
  \abstract
   {The flat spectrum radio quasar (FSRQ) OP~313 was discovered in December 2023 in very-high-energy $\gamma$ rays above 100~GeV, enabling for the first time a complete broadband characterization of its emission. However, the lack of updated {measurements} of its accretion disc, broad line region, and dusty torus {hampers} a detailed interpretation of the role of {accretion} in the observed $\gamma$-ray {production}.}
   {We intended to characterize, {during high-activity states}, the brightness of the external photon fields {contributing} to the infrared-to-ultraviolet emission---namely the dusty torus, broad line region, and accretion disc---as well as {to investigate potential variability and blurring effects} on the optical broad emission lines. We also aimed to {constrain} the particle population responsible for the continuum non-thermal synchrotron emission.}
   {We present new spectroscopic observations of OP~313 with the NOT and TNG telescopes in order to characterize its optical spectrum  and  variability with respect to archival observations from SDSS performed {during a} low emission state. We measured the luminosity of  different broad emission lines, evaluating possible changes in the broad line region luminosity. These measurements also enabled an updated characterization of the broad line region, accretion disc, and dusty torus properties.}
   {We measured the Mg {\tiny II} emission line, detectable in six of the seven spectra, with an average flux of $F_{\mathrm{Mg \ {\tiny II}}} = (0.85 \pm 0.11)\times 10^{-14}$~erg~cm$^{-2}$~s$^{-1}$. 
   {Its equivalent width and luminosity are consistent with a constant emission line with a variable non-thermal continuum that buries other lines ($H\delta$ and  $H\gamma$). From the stable Mg {\tiny II} line we derived a constant luminosity of the thermal components, finding  $\log(L_{\mathrm{BLR}} \ \mathrm{[erg \ s^{-1}]}) = 44.91 \pm 0.19$, $\log(L_{\mathrm{disc}} \ \mathrm{[erg \ s^{-1}]}) = 45.91 \pm 0.19$, and $\log(L_{\mathrm{torus}} \ \mathrm{[erg \ s^{-1}]}) = 44.70 \pm 0.16$, and an estimated black hole mass of $\log(M_{BH}/M_{\odot})=8.36 \pm 0.18$.} These estimates are in line with with those derived from the C~{\tiny III}] emission line, detected in five of the seven spectra. The characteristics derived from the emission line and the indicator of the accretion rate from the disc/Eddington luminosity ratio $\lambda =L_{AD}/L_{Edd} = 0.23 \pm 0.10$, along with a high Compton dominance observed by previous studies, favour a FSRQ-like nature of OP~313, contrary to the argued changing-look nature of the source. Under this scenario, the $\gamma$-ray variability is fully attributed to changes in the particle population, rather than variations in the actual thermal structure.}
   {}


   \maketitle
%

\section{Introduction}
The optical emission of blazars --- active galactic nuclei (AGNs) with highly luminous jets of ultra-relativistic particles closely aligned with our line of sight --- is typically dominated by a very bright {and variable} non-thermal synchrotron continuum \citep[see e.g.][]{koenigl1981}. {This is especially the case of the BL Lacertae blazar subclass, where no strong thermal emission is expected, as they are often assumed to have a weak accretion disc \citep{wang2002}, and similarly faint to non-existent dusty torii and gas clouds producing broad emission lines via photoionization around the central black hole \citep{cao2004,plotkin2012}. As a result, their spectra are almost always featureless or, at most, have very faint emission lines only visible during low emission states.} 

{In contrast, flat spectrum radio quasars (FSRQs) have more efficient accretion flows \citep{jolley2009}, which indirectly results in efficient reprocessing of the emission by the dusty torus into the infrared band \citep{haas1998} and into bright broad emission lines in the optical band \citep{padovani1992} by the broad line region (BLR).}
While the relative contribution of these thermal components is heavily dependent on the flux level of the variable non-thermal synchrotron continuum, they have been observed to be even brighter than the jet {in many FSRQs} \citep[see e.g.][]{raiteri2014}. Therefore, disentangling these contributions is key to {understanding} the optical emission of blazars, as they teach us about the accretion rate, the density and clumpy nature of the dusty structures, the kinematic of the gas clouds, and the reprocessing efficiency of both components.
{These are all urging questions in the study of FSRQs, playing} a pivotal role in the interpretation of the high-energy ($E>100$~MeV) and very-high-energy ($E>100$~GeV) $\gamma$-ray emission observed from them. Under a leptonic scenario, this emission is explained in the framework of the inverse-Compton (IC) scattering of low-energy photons with the relativistic electrons from the jet. When the interacting photons are the ones produced in the jet via synchrotron radiation, this process is often known as synchrotron self-Compton (SSC) scattering \citep{maraschi1992}. However, when there are external photon fields that provide low-energy seed photons to the jet that then interact via IC with the relativistic electrons, the resulting scattering is known as external Compton \citep[EC, see][]{dermer1993}. For FSRQs, the BLR, accretion disc, and dusty torus  are all possible sources of bright external photon fields to the IC process that often result in a strong EC contribution to the $\gamma$-ray emission that dominates over SSC{, and consequently lead to significant radiative cooling of the relativistic electrons and efficient pair production at the highest energies, shaping the observed spectral energy distribution and limiting the maximum energies they can attain.} 

OP~313 is a FSRQ located at redshift $z=0.997$ \citep{schneider2010}. It has been regularly detected by the \textit{Fermi}-LAT telescope in high-energy $\gamma$ rays since 2019. Starting in November 2023, the source has been in a period of violent variability and recurrent outbursts of very bright $\gamma$-ray emission \citep{bartolini2023}, making it one of the most luminous $\gamma$-ray sources ever detected and leading to its first detection in the very-high-energy $\gamma$-ray domain by LST-1 \citep{cortina2023}, being only the tenth FSRQ ever detected at such high energies, and the most distant of all of them. This increased activity was also accompanied by a high optical flux \citep{otero2023} due to an enhancement of the synchrotron continuum, {changing the relative contribution of the different radiation fields.} 
Through optical spectroscopic observations in different phases of the flare, we aimed to constrain the variability of the different components and ultimately: i) measure the luminosity of the accretion disc, BLR, and dusty torus
; ii) study potential BLR luminosity variability that could point towards changes in the accretion disc during the flare; and iii) constrain the nature of the jet particle distribution and its non-thermal energy spectrum through a high precision spectroscopic measurement of the photon field in the synchrotron component.

The paper is structured as follows: in Sect.~\ref{sec2} we detail the observations used in this work and the data reduction; in Sect.~\ref{sec3} we present the identification and modelling of emission lines and the evaluation of possible variations between the different epochs; in Sect.~\ref{sec4} we perform the characterization of the thermal components of the AGN based on the luminosity of the emission lines; and in Sect.~\ref{sec5} we evaluate the variability of the synchrotron continuum and the underlying electron population. Finally, in Sects.~\ref{sec6} and \ref{sec7}, we conclude with a discussion and final remarks of the main results of this work.


\section{Observations and data reduction}\label{sec2}
Spectroscopic observations were performed during four nights between May and July 2024 in order to characterize the optical emission of the source and relative contribution of the non-thermal and thermal components. These observations are coincident with a period of large $\gamma$-ray and optical variability of the source after a long-lasting flaring state that started in November 2023. These observations were performed with the 2.5-m Nordic Optical Telescope (NOT) and the 3.58-m Telescopio Nazionale Galileo (TNG), located at the Roque de los Muchachos Observatory in the Canary island of La Palma. 

The NOT spectra were taken using the ALFOSC\footnote{\url{https://www.not.iac.es/instruments/alfosc/}} (Alhambra Faint Object Spectrograph and Camera) instrument. The observations took place during the nights of May 1 and June 18, 2024. In order to cover a wide spectral range, two spectra were taken each night (see Table~\ref{tab:obslog} the setup details), each one covering the blue and red parts of the optical spectrum for a total coverage of $3200-10150$~\AA.  Spectra of the spectrophotometric standard HD~93521 \citep{oke1990} were taken for the flux calibration.

A similar strategy was followed with the TNG on the night of May 28, 2024, using the DOLORES\footnote{\url{https://www.tng.iac.es/instruments/lrs/}} (Device Optimized for the LOw RESolution) spectrograph. The blue and red parts of the spectrum were covered separately using the LR-B and LR-R grisms provided by the instrument. These grisms provide a wavelength coverage of $3000-8430$~\AA~and $4470-10070$~\AA.
An observation of the standard spectrophotometric standard Feige~34 performed on the same night was used for the flux calibration \citep{oke1990}. 

An additional observation also took place on the night of July~3, 2024, again with the TNG telescope, in this case focusing only on the blue part of the spectrum using DOLORES and grism LR-B. The standard spectrophotometric star HZ~44 \citep{oke1990} was observed for an accurate flux calibration of the spectra. Photometric observations were also performed almost in parallel to the spectroscopic data acquisition.
Details about all the observations are listed in Table \ref{tab:obslog}.

\begin{figure*}[h!]
    \centering
    \includegraphics[width=\linewidth]{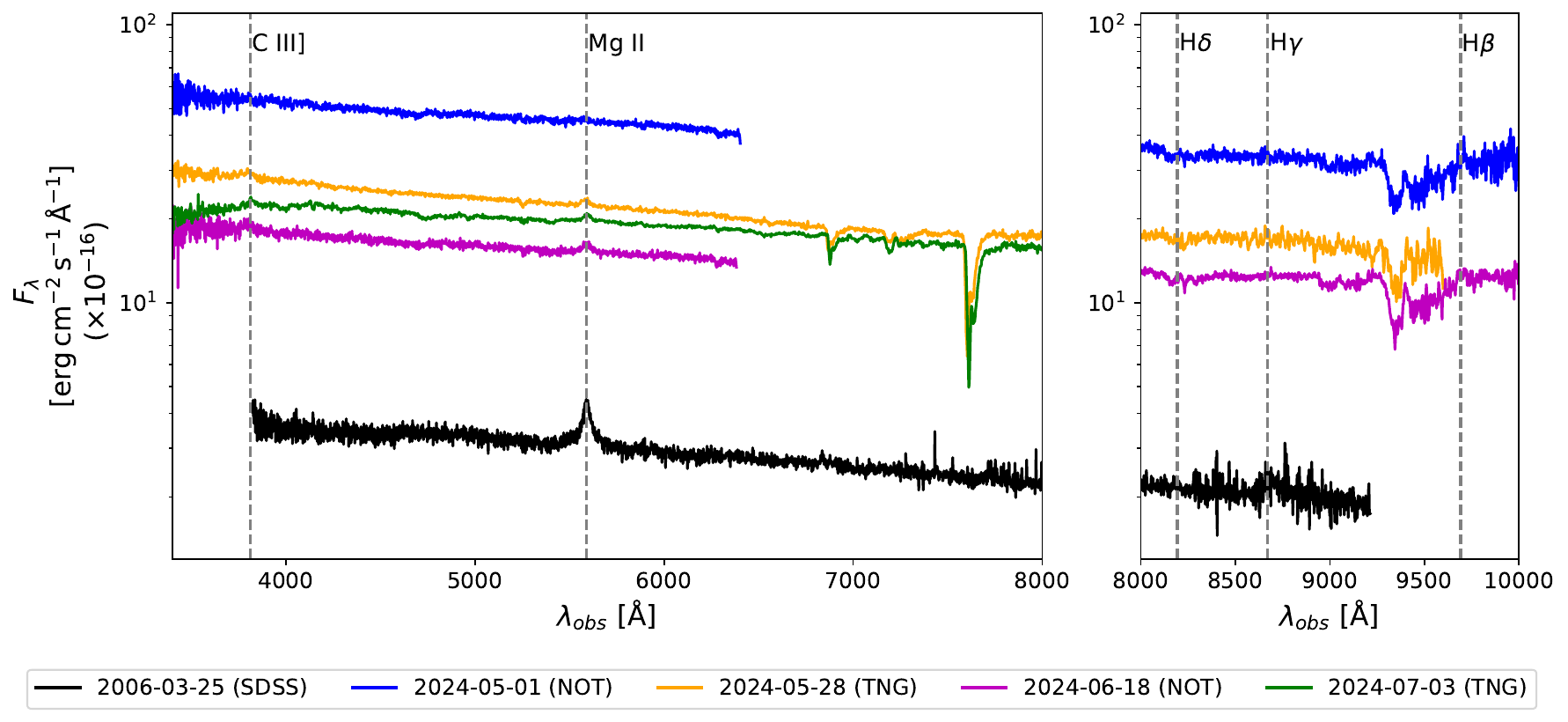}
    \caption{Optical spectra of OP~313 observed with NOT and TNG telescopes. \textit{Left:} Blue spectra. \textit{Right:} Red spectra. The NOT spectrum from May 1 and June 18 are represented in blue and magenta, respectively. The TNG spectra from May 28 and July 3 are represented in orange and green, respectively. The TNG spectra represented here correspond to the co-added spectra of both exposures of each night. The SDSS spectrum taken on March 25, 2006, is also shown in black. The positions of the brightest emission lines reported in \cite{vandenberk2001} with a relative flux to the $Ly\alpha$ line $100 \times F/F_{Ly\alpha} \geq 1$ and within the spectral range of the observations presented in this work are highlighted with vertical dashed grey lines and indicated in the figure.}
    \label{fig:optical_spectra}
\end{figure*}

\begin{table*}
    \centering
    \caption{Summary of the configuration of all OP 313 spectral observations performed in 2024 with the NOT and TNG telescopes.}
    \begin{tabular}{cccccccc}
      \hline
         \multirow{2}{*}{Observation Date} & \multirow{2}{*}{MJD} &  \multirow{2}{*}{Instrument} &  \multirow{2}{*}{Grism/Slit} & Spectral range &  Dispersion &  \multirow{2}{*}{R} & $t_{exp}$\\
       &  &  &  & [\AA] & [\AA/pixel] & & [s]\\\hline
        
        \multirow{2}{*}{May 1, 2024}& \multirow{2}{*}{60431} & NOT/ALFOSC & \#14/1.8\arcsec & $3200-6380$ & 1.603 & 550 & 1200  \\
                &   & NOT/ALFOSC &  \#20/1.8\arcsec & $5650-10150$  & 2.196 & 560 & 500 \\ \hline
        \multirow{2}{*}{May 28, 2024} & \multirow{2}{*}{60458} & TNG/DOLORES & LR-B/1.5\arcsec & $3000-8430$ & 2.738 & 390 & $2\times600$ \\
                &    & TNG/DOLORES & LR-R/1.5\arcsec & $4470-10070$  & 2.705 & 475 & $2\times450$ \\ \hline
        \multirow{2}{*}{June 18, 2024} & \multirow{2}{*}{60479} & NOT/ALFOSC & \#14/1.8\arcsec & $3200-6380$  & 1.603 & 550 & 1200 \\
                &    & NOT/ALFOSC & \#20/1.8\arcsec & $5650-10150$  & 2.196 & 560 & 500   \\ \hline
        July 3, 2024 & 60494 & TNG/DOLORES & LR-B/1.5\arcsec &  $3000-8430$ & 2.738  & 390 & $2\times1000$ \\ \hline
    \end{tabular}
    \label{tab:obslog}
\end{table*}

The data reduction  of the NOT and TNG spectral data was performed using the \texttt{PYPEIT} software package \citep{Prochaska2020}. The standard procedure is followed to subtract the bias and to correct by flat-field. The 1D-spectra are extracted following the optimal  extraction algorithm \citep{Horne1986}, after a local sky spectrum was subtracted.
The resulting spectra are shown in Fig.~\ref{fig:optical_spectra}.
The emission was corrected from air mass extinction.
The absolute flux level is finally checked using Sloan $g$, $r$ and $i$ photometry obtained on the same nights with the imaging mode of ALFOSC or DOLORES, right before the spectral observations.

In addition, we have retrieved an archival optical spectrum of OP~313 publicly available at the Sloan Digital Sky Survey (SDSS) database\footnote{\url{https://www.sdss.org}} from the SDSS legacy program \citep{york2000,abdurrouf2022}. A detailed description of the SDSS legacy program, target selection, instrumentation, and data analysis can be found in \cite{richard2002,smee2013} and references therein. This spectrum, taken on March 25, 2006, covers a wavelength range between $\sim$3700~\AA~and $\sim$9300\AA. This spectrum is also shown in Fig.~\ref{fig:optical_spectra}. We represent the spectra in the wavelength range between 3400~\AA~and 10000~\AA, since outside this range they are dominated by noise and no features are observed.

\section{Line variability and blurring}\label{sec3}
We have identified a set of broad emission lines within the spectral range of the different optical spectra shown in Fig.~\ref{fig:optical_spectra}. We modelled the profile of the detected lines in order to characterize the flux and luminosity  and evaluate possible temporal variability of the line emission. We performed the line fit for those lines  detected over the non-thermal emission and background noise, i.e. those for which the equivalent width is larger than the minimum measurable EW$_{\text{min}}$.
For this, we followed the procedure applied by \cite{becerra2021} and detailed in Appendix~\ref{sec:A1}. Additionally, to guarantee that the emission lines are well resolved from the continuum noise, we considered only those for which more than two-thirds of the pixels within the wavelength window considered for the line characterization are above the continuum.
This leaves us with the Mg {\tiny II} and C {\tiny III}] emission lines as the only ones that are consistently resolved in a significant way, with the exception of the spectrum from May~1, 2024, for which they were not detected. The C {\tiny III}] emission line is significantly resolved for the spectra taken on the nights of May 28, June 18, and July 3. We note that it was not detected and characterized in the SDSS spectrum as it falls right in the edge of the spectral range. Other bright emission lines within the spectral range studied here such as $H\delta$ or $H\gamma$, whose positions are highlighted in Fig.~\ref{fig:optical_spectra}, are not significantly resolved from the continuum.

\begin{figure}
    \centering
    \includegraphics[width=1\linewidth]{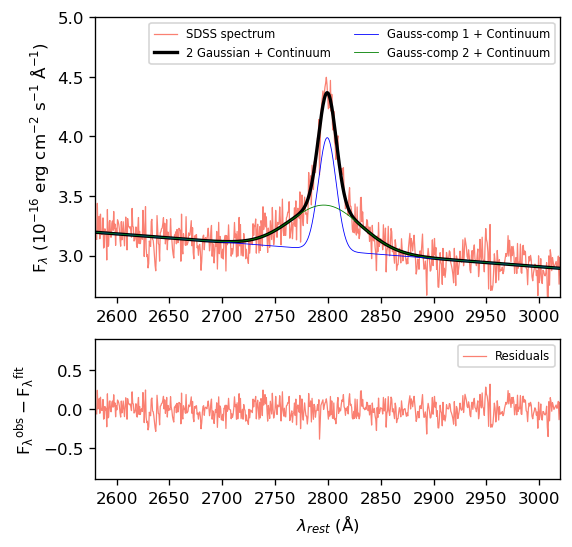}
    \caption{Modelling of the Mg {\tiny II} line profile for the spectrum taken by the SDSS with two Gaussian components plus a linear function to model the synchrotron continuum. The bottom panel shows the residuals of the fit.}
    \label{fig:MgII_line}
\end{figure}

The fit of the emission line profile was performed as a composite of Gaussian functions plus a linear function that represents the continuum synchrotron emission. {The choice of a linear function over the  power law commonly used to represent blazars' synchrotron continuum is due to the narrow spectral range fitted for the line characterization. In this short wavelength range, both functions provided an accurate fit of the non-thermal continuum. Hence, the simple linear function was used.} For the spectra taken on May 28 and June 18, 2024, we observed that the broad component of the Mg {\tiny II} line is covered by the continuum emission. Therefore, these three spectra were fitted with a single Gaussian component. All the resolved C {\tiny III}] lines were modelled with a single Gaussian function. An example is shown in Fig.~\ref{fig:MgII_line}, where the characterization of the Mg {\tiny II} emission line is presented for the spectrum taken by the SDSS. The figures showing the fits for the remaining spectra {are presented} in Appendix~\ref{sec:A2} (see {also} Figs.~\ref{fig:MgII_line_all} and~\ref{fig:CIII_line_all} for the Mg {\tiny II} and C {\tiny III}] lines, respectively). Then, we characterized the emission line, i.e. its centroid, EW, and full width half maximum (FWHM) using the \textsc{python} package \texttt{specutils} \citep{specutils2019}. Finally, we also calculated the flux and luminosity of the emission line. The conversion from flux to luminosity units was performed assuming a flat $\Lambda$CDM cosmological model with parameters $H_0=69.6$~km~s$^{-1}$, $\Omega_m = 0.286$, and $\Omega_{\Lambda}=0.714$ \citep{bennett2014}. This combination of parameters yields a luminosity distance $d_L = 6676.2 \ \text{MPc} = 2.06 \times 10^{28} \ \text{cm}$ at $z=0.997$. The derived parameters for each emission line characterized are summarised in Table \ref{tab:emission_lines}.

\begin{table*}
\centering
\caption{Measurement of Mg {\tiny II} and C {\tiny III}] optical emission lines from OP 313 for the different spectra obtained for this work and the archival spectrum from the SDSS database.}
\label{tab:emission_lines}
\resizebox{\textwidth}{!}{%
\begin{tabular}{cccccccc}
\hline
 \multirow{2}{*}{Line} & \multirow{2}{*}{Date} & \multirow{2}{*}{Instrument} & $\lambda_{cent}$$^{a}$ & FWHM & EW$_{\text{obs}}$ & $F_{\mathrm{line}}$$^{b}$ & $\log (L_{\mathrm{line}})$ \\ 
&    &    & [\AA]  & [km~s$^{-1}$] & [\AA] & [$\times 10^{-14}$~erg~cm$^{-2}$~s$^{-1}$] & [erg~s$^{-1}$] \\ \hline
\multirow{7}{*}{Mg {\tiny II}} &  2006-03-25 & SDSS & $2799.4 \pm 0.4$ & $2680 \pm 147$ & $-33.52 \pm 1.25$ & $ 1.02 \pm 0.11$ & $43.74 \pm 0.05$ \\ 
&  2024-05-01 & NOT/ALFOSC  & -- & -- & $-0.88 \pm 0.48$ & -- & -- \\ 
&  2024-05-28 & TNG/DOLORES & $2797.0 \pm 1.1$  & $2371 \pm 304$ & $-2.77 \pm 0.34$ & $0.56 \pm 0.12$ & $43.47 \pm 0.09$ \\ 
&  2024-05-28 & TNG/DOLORES & $2795.3 \pm 2.0$  & $2805 \pm 743$ & $-3.33 \pm 0.35$ & $0.63 \pm 0.14$ & $43.53 \pm 0.10$ \\ 
&  2024-06-18 & NOT/ALFOSC  & $2800.0 \pm 1.3$  & $3768 \pm 543$ & $-6.58 \pm 0.34$ & $0.82 \pm 0.11$ & $43.64 \pm 0.06$ \\ 
&  2024-07-03 & TNG/DOLORES & $2799.5 \pm 1.0$  & $2163 \pm 375$ & $-7.31 \pm 0.36$ & $ 1.45 \pm 0.23$ & $43.89 \pm 0.07$ \\ 
&  2024-07-03 & TNG/DOLORES & $2801.4 \pm 1.8$  & $3125 \pm 544$ & $-7.17 \pm 0.42$ & $ 1.40 \pm 0.24$ & $43.88 \pm 0.08$ \\ \hline
\multirow{4}{*}{C {\tiny III}]}   &  2024-05-01  &  NOT/ALFOSC   & -- & -- & $-0.49 \pm 0.42$ & -- & -- \\ 
     & 2024-05-28  &  TNG/DOLORES   & $1901.6 \pm 2.2$ & $4236 \pm 713$ & $-3.73 \pm 0.49$ & $0.94 \pm 0.20$ & $43.70 \pm 0.09$ \\ 
     &  2024-05-28  &  TNG/DOLORES   & $1895.5 \pm 3.2$ & $8395 \pm 4564$ & $-4.11 \pm 0.50$ & $1.35 \pm 0.75$ & $43.86 \pm 0.24$ \\
     &  2024-06-18  &  NOT/ALFOSC   & $1902.1 \pm 2.7$ & $4140 \pm 1184$ & $-4.03 \pm 0.52$ & $0.78 \pm 0.18$ &  $43.62 \pm 0.10$ \\ 
     &  2024-07-03  &  TNG/DOLORES   & $1911.2 \pm 2.9$ & $6964 \pm 2745$ & $-5.17 \pm 0.42$ & $1.27 \pm 0.41$ & $43.83 \pm 0.14$ \\
     &  2024-07-03  &  TNG/DOLORES   & $1911.4 \pm 1.7$ & $9261 \pm 2934$ & $-4.41 \pm 0.50$ & $1.33 \pm 0.39$ & $43.85 \pm 0.13$ \\ \hline
\end{tabular}
}\\
{\flushleft
\vspace{-0.25cm}
{\small \textit{Notes.} $^a$Restframe wavelength. $^b$We have considered a 10\% systematic uncertainty associated to the photometric calibration.}\\
}
\end{table*}

As evidenced by the EW values, we are only able to resolve significantly the broad component of the Mg {\tiny II} line during the low state (SDSS spectrum taken on 2006) and during the lowest states of the observations performed in 2024, {that is}, for the spectra taken on July 3, and marginally (at a $1\sigma$ level) on June 18. On the other hand, on May 1 and May 28, the dominance of the continuum, with an increase of more than a factor 10 (see Fig.~\ref{fig:optical_spectra}) hinders the broad Mg {\tiny II} component, resulting in $\mathrm{|EW|}<5$~\AA, completely covering the line on the former night. This clearly affects as well the measured integrated flux of the emission line and therefore its luminosity. The integrated Mg {\tiny II} flux shows variations of $40$\% with respect to the value derived from the SDSS spectrum. Nevertheless, for the spectra during May 28 we observe a lower flux as a result of resolving only the narrow component, whereas the flux of the other spectra is consistent within errors. The average values of the Mg {\tiny II} line flux and luminosity are $F_{\mathrm{Mg \ {\tiny II}}} = (0.85 \pm 0.11) \times 10^{-14}$~erg~cm$^{-2}$~s$^{-1}$ and $\log(L_{\mathrm{Mg \ {\tiny II}}} \ \mathrm{[erg \ s^{-1}]}) = 43.72 \pm 0.07$. These values are consistent with previous estimations of the Mg~{\tiny II} line flux for OP~313 \citep[see e.g.][]{shaw2012,pandey2025}.

The FWHM of this line takes values between $\sim$2200 and $\sim$2800~km~s$^{-1}$ and an average of $\mathrm{FWHM} = 2654 \pm 350$~km~s$^{-1}$, with two spectra showing somewhat higher values. For the second exposure performed on July 3, we measure a FWHM of 3125~km~s$^{-1}$, with however a rather large uncertainty, making it compatible with the average value. On the other hand, the spectrum taken on June 18 shows the highest FWHM value, 3768~km~s$^{-1}$. This is compatible with an optical spectrum taken independently at Calar Alto Observatory on the same night and reported in \cite{pandey2025}, where they measure a consistent value, higher than that from the SDSS spectrum.

We have also characterized the C {\tiny III}] emission line for the five spectra in which this line was significantly resolved, as specified above. The average C {\tiny III}] flux and luminosity values obtained are $F_{\mathrm{C {\tiny \ III}]}}= (1.00 \pm 0.10) \times 10^{-14}$~erg~cm$^{-2}$~s$^{-1}$ and $\log(L_{\mathrm{C \ {\tiny III}]}} \ \mathrm{[erg \ s^{-1}]}) = 43.77 \pm 0.10$. The values for each spectrum are reported in Table~\ref{tab:emission_lines}. The FWHM adopts systematically higher values in comparison to those from the Mg {\tiny II} line. Nevertheless, we note that this line lies in a noisier part of the spectrum, as reflected by the lower signal-to-noise (S/N) shown in Fig.~\ref{fig:CIII_EWmin} in comparison to that of the Mg {\tiny II} reported in Fig.~\ref{fig:MgII_EWmin}. The average value is $\mathrm{FWHM} = 6599 \pm 940$~km~s$^{-1}$. \cite{finke2016} reports that both the Mg {\tiny II} and C \ {\tiny III}] emission lines contribute similarly to the total luminosity of the BLR (1.7 and 1.8 times the luminosity of the $H\beta$ line, respectively). Therefore, their very similar flux and luminosity are consistent with this assumption.

\begin{figure}
    \centering
    \includegraphics[width=1\linewidth]{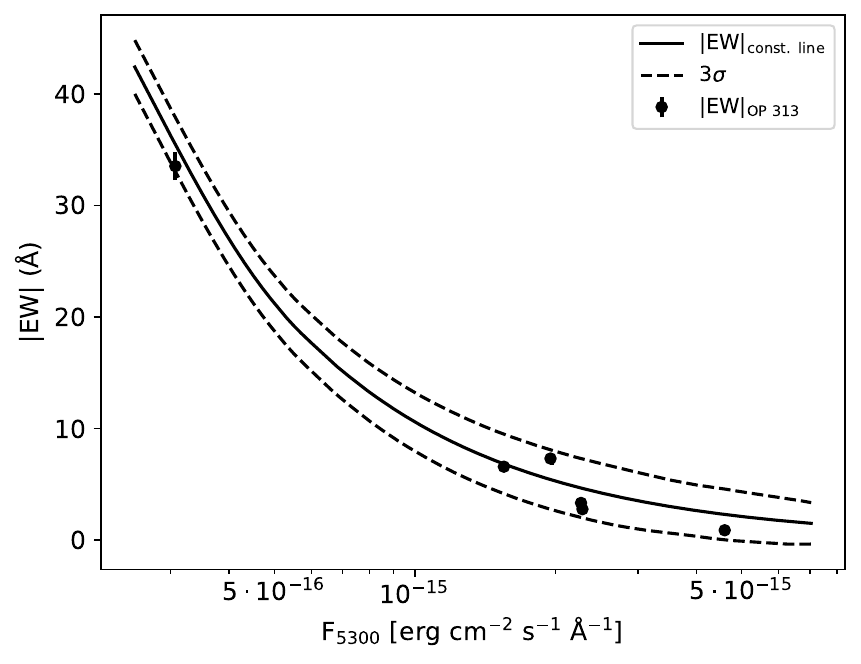}
    \caption{EW measured for each spectrum obtained for OP 313 with respect to the relation between the EW of a constant Mg {\tiny II} emission line and the continuum flux evaluated at 5300~\AA. The dashed lines represent the 2$\sigma$ confidence level of the derived relation.}
    \label{fig:EW_continuum}
\end{figure}

In order to further evaluate possible variations in the emission lines we have compared the EW  measured in each spectrum with the expected relation between EW and continuum flux following the methodology presented by \cite{isler2013}. We applied this methodology to the six spectra in which we significantly resolved the Mg {\tiny II} line. For this, we first modelled the Mg {\tiny II} and measured its EW and the continuum at 5300~\AA~for the SDSS spectrum. Then, assuming a constant Mg {\tiny II} line, we increased the flux of the modelled continuum component in fine steps, re-evaluating in each step the expected EW for each new continuum flux level. Finally, through Monte Carlo (MC) simulations we evaluated the 2$\sigma$ confidence level of the obtained EW-continuum relation. In Fig.~\ref{fig:EW_continuum} we show the EW measured for each spectrum analysed here with respect to the expected EW-continuum relation. The EW measurements of all six spectra are well contained within the 3$\sigma$ confidence interval of this relation. Therefore, the results obtained here are consistent with a constant emission line profile along the different periods studied here.

\begin{figure}
    \centering
    \includegraphics[width=1\linewidth]{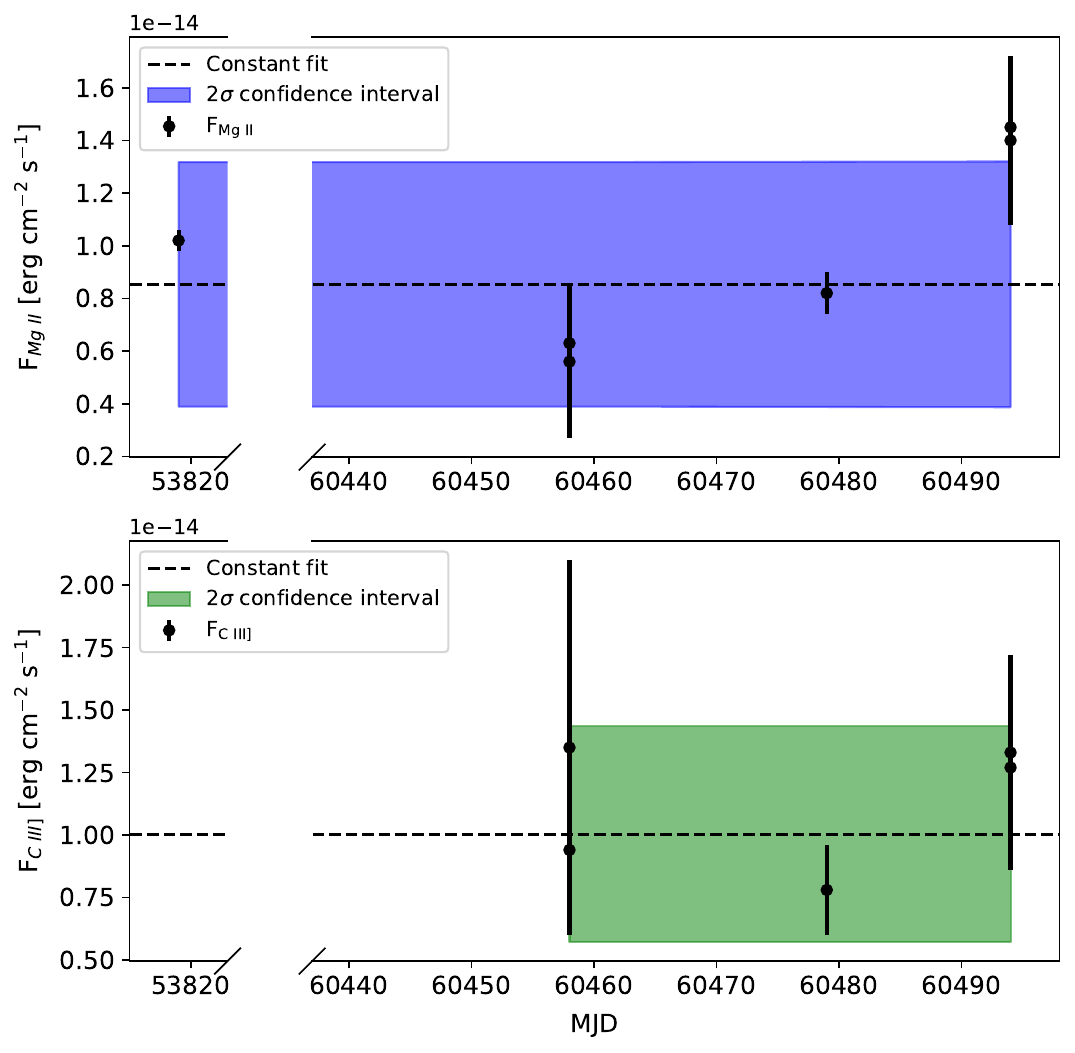}
    \caption{Constant fit to the measured flux of the Mg {\tiny II} and C {\tiny III}] emission lines over time for the different spectra analysed here. \textit{Top}: Mg {\tiny II} line. \textit{Bottom}: C {\tiny III}] line. The black points correspond to the observational measurements. The black dashed line is the best fit constant value obtained and the blue and green contours represents the 2$\sigma$ confidence interval of the fit in each case.}
    \label{fig:MgII_CIII_flux_vs_time}
\end{figure}

Alternatively, we tested a second approach proposed by \cite{chavushyan2020}, based on a constant fit of the measured emission line flux over time. Again, we applied it to the Mg~{\tiny II} emission line for those six spectra in which it is resolved from the continuum. In the top panel of Fig.~\ref{fig:MgII_CIII_flux_vs_time} we show the fit to a constant luminosity for the Mg {\tiny II} lines identified in the different spectra. We observe that all measurements are consistent with the expected values from a constant line within errors at a 2$\sigma$ confidence level, even for the spectra where the continuum level outshined the emission line almost completely and where the broad component was not visible. This can also be extended to the fluxes measured for the C {\tiny III}] emission line (see bottom panel of Fig.~\ref{fig:MgII_CIII_flux_vs_time}), in this case even within a 1$\sigma$ level, since the measurements are in a more noisy region of the spectrum and therefore have larger uncertainties (see Table~\ref{tab:emission_lines}).

\section{Constrains on the thermal emission}\label{sec4}
Based on the estimations of the luminosity of the different lines observable in the optical spectra, we can obtain estimates for the BLR, accretion disc, and dusty torus luminosities. For this we have used the measurements of the Mg {\tiny II} line as the most prominent one observed in the spectra, using the luminosity values reported in Table~\ref{tab:emission_lines} and the relative contributions from each emission line to the emission of the BLR from \cite{finke2016} in order to scale the observed line luminosity to the total BLR luminosity. As a crosscheck, we have compared the results for the Mg {\tiny II} emission line with those obtained with the line contribution derived by \cite{francis1991} of 34/556 assuming that the Ly$\alpha$ emission line has a contribution of 100, obtaining consistent values within errors \citep[see also][]{ghisellini2015}.  

\begin{table*}
\centering
\caption{Characterization of the accretion disc, BLR, and dusty torus, and the black hole mass of OP 313 as derived from the different measurements of the Mg {\tiny II} and C {\tiny III}] emission lines.}
\label{tab:thermal_components}
\resizebox{\textwidth}{!}{%
\begin{tabular}{ccccccccc}
\hline
 \multirow{2}{*}{Date} & \multirow{2}{*}{Line} & \multirow{2}{*}{Instrument} & $\log(L_{\text{disc}})$ & $\log(L_{\text{BLR}})$ & $r_{\text{BLR}}$ & $\log(L_{\text{torus}})$ & $r_{\text{torus}}$ & \multirow{2}{*}{$\log{(M_{BH}/M_{\odot}})$} \\ 
   &   &    & [erg~s$^{-1}$]  & [erg~s$^{-1}$] & [$\times$10$^{17}$~cm] & [erg~s$^{-1}$] & [$\times$10$^{18}$~cm] & \\ \hline
 2006-03-25  & Mg {\tiny II} &  SDSS  & $45.96 \pm 0.05$ & $44.96 \pm 0.05$ & $3.00 \pm 0.16$ & $44.74 \pm 0.05$ & $7.52 \pm 0.40$ & $8.39 \pm 0.14$\\ \hline
 \multirow{4}{*}{2024-05-28} & Mg {\tiny II} &  \multirow{4}{*}{TNG/DOLORES}  & $45.61 \pm 0.08$ & $44.61 \pm 0.08$ & $2.21 \pm 0.17$ & $44.47 \pm 0.08$ & $5.53 \pm 0.42$ & $8.12 \pm 0.15$  \\ 
  &  Mg {\tiny II} & & $45.70 \pm 0.10$ & $44.70 \pm 0.10$ & $2.37 \pm 0.25$ & $44.53 \pm 0.10$ & $5.92 \pm 0.64$ &  $8.30 \pm 0.29$ \\ 
  &  C {\tiny III}] & & $45.90 \pm 0.09$ & $44.90 \pm 0.09$ & $2.81 \pm 0.30$ & $44.68 \pm 0.09$ & $7.03 \pm 0.76$ & --  \\ 
  &  C {\tiny III}] & & $46.05 \pm 0.24$ & $45.05 \pm 0.24$ & $3.36 \pm 0.94$ & $44.83 \pm 0.24$ & $8.41 \pm 2.35$ & --  \\ \hline
  \multirow{2}{*}{2024-06-18} & Mg {\tiny II}  &  \multirow{2}{*}{NOT/ALFOSC}  & $45.86 \pm 0.06$ & $44.86 \pm 0.06$ & $2.70 \pm 0.20$ & $44.64 \pm 0.06$ & $6.75 \pm 0.49$ &  $8.63 \pm 0.18$ \\ 
  & C {\tiny III}] &    & $45.81 \pm 0.10$ & $44.81 \pm 0.10$ & $2.55 \pm 0.30$ & $44.59 \pm 0.10$ & $6.37 \pm 0.76$ & --  \\    \hline
  \multirow{4}{*}{2024-07-03}  &  Mg {\tiny II}  &  \multirow{4}{*}{TNG/DOLORES}  & $46.11 \pm 0.09$ & $45.11 \pm 0.09$ &  $3.59 \pm 0.36$ & $44.89 \pm 0.09$ & $8.97 \pm 0.90$ & $8.30 \pm 0.52$  \\ 
   &   Mg {\tiny II}   &   & $46.10 \pm 0.07$ & $45.10 \pm 0.07$ &  $3.53 \pm 0.27$ & $44.87 \pm 0.07$ & $8.83 \pm 0.68$ &  $8.61 \pm 0.37$   \\
   & C {\tiny III}]     &   & $46.03 \pm 0.13$ & $45.03 \pm 0.13$ & $3.26 \pm 0.50$ & $44.80 \pm 0.13$ & $8.15 \pm 1.25$ & --  \\ 
   & C {\tiny III}]   &  & $46.05 \pm 0.12$ & $46.05 \pm 0.12$ & $3.34 \pm 0.48$ & $44.83 \pm 0.13$ & $8.35 \pm 1.21$ & --  \\  \hline
\end{tabular}
}\\
{\flushleft
\vspace{-0.25cm}
{\small \textit{Notes.} Uncertainties correspond to statistical errors plus the considered 10\% systematic associated with the photometric calibration.}\\
}
\end{table*}

Then, the BLR luminosity can be related to the accretion disc luminosity as
\begin{equation}
L_{\text{disc}} \approx \frac{1}{f_{\text{cov}_{\text{BLR}}}} \times L_{\text{BLR}},
\label{eq:accretion_disc_luminosity}
\end{equation}
where $f_{\text{cov}_{\text{BLR}}}$ is the covering factor of the accretion disc. We assumed a standard value of 0.1 for $f_{\text{cov}_{\text{BLR}}}$ \citep[see e.g.][]{becerra2021}.
The radius of the BLR can be inferred as well from the estimate of the accretion disc luminosity assuming the BLR as an homogeneous shell as \citep{ghisellini2009}
\begin{equation}
r_{\text{BLR}} \approx 10^{17} \left[ \frac{L_{\text{disc}}}{10^{45}~\text{erg~s}^{-1}} \right]^{0.5} \ \text{cm}.
\label{eq:blr_radius}
\end{equation}  

Analogously, the relation between BLR and dusty torus luminosities can be expressed as
\begin{equation}
L_{\text{torus}} \approx f_{\text{cov}_{\text{torus}}} \times L_{\text{BLR}},
\label{eq:torus_luminosity}
\end{equation}
being $f_{\text{cov}_{\text{torus}}}$ the covering factor of the torus. Typically, values between 0.5 and 0.7 are used in the literature for $f_{\text{cov}_{\text{torus}}}$. Here we adopted an intermediate value of 0.6, following the prescription from \cite{becerra2021}. 
Moreover, the size of the dusty torus can be derived from the disc luminosity following the prescription of \cite{ghisellini2009} under the assumption of a homogeneous dusty torus as
\begin{equation}
r_{\text{torus}} \approx 2.5 \times 10^{18} \left[ \frac{L_{\text{disc}}}{10^{45}~\text{erg~s}^{-1}} \right]^{0.5} \ \text{cm},
\label{eq:torus_radius}
\end{equation}
being $r_{\text{torus}}$ the radius of the torus. While we note that the morphology of the torus can be more complicated that the one assumed here, possibly with clumpy and inhomogeneous structures \citep{nenkova2008,lopez2025}, this estimate provides a reasonable approximation to the nature of the torus, providing a simplified picture of the morphology of the inner parts of the AGN. We summarise in Table~\ref{tab:thermal_components} the results and estimates of the accretion disc, BLR, and dusty torus obtained for OP~313 based on the different broad lines observed in the optical spectra.

Considering all the spectra analysed here, we have estimated an average luminosity of the BLR, accretion disc, and dusty torus of $\log(L_{\mathrm{BLR}} \ \mathrm{[erg \ s^{-1}]}) = 44.91 \pm 0.19$, $\log(L_{\mathrm{disc}} \ \mathrm{[erg \ s^{-1}]}) = 45.91 \pm 0.19$, and $\log(L_{\mathrm{torus}} \ \mathrm{[erg \ s^{-1}]}) = 44.70 \pm 0.16$, respectively. Moreover, we also estimated the radius of the BLR and torus, showing average values of $r_{\mathrm{BLR}} = (2.76 \pm 0.53) \times 10^{17}$~cm and $r_{\mathrm{torus}} = (6.89 \pm 1.32) \times 10^{18}$~cm. This values are consistent with the estimation reported for the accretion disc by \cite{ghisellini2015}. On the other hand, the accretion disc luminosity is lower than that reported by \cite{pandey2025} between $46.42 \pm 0.26$ and  $46.53 \pm 0.26$, calculated however through a different relation with the Mg {\tiny II} line luminosity. It is also worth mentioning that the values reported here depend on the covering factors of the disc and torus. Therefore, despite both values being comparable, assuming a higher covering factor instead of the rather low value of 0.1 used in our analysis would overcome the observed difference. 

As a comparison, we have also calculated the luminosity of the thermal components from the characterization of the C {\tiny III}] {under the same assumptions of covering factors for the BLR ($f_{\text{cov}_{\text{BLR}}} = 0.1$) and torus ($f_{\text{cov}_{\text{torus}}}=0.6$) as for the Mg {\tiny II} line}. These values are also reported in Table~\ref{tab:thermal_components}. We obtained consistent values with those reported above, estimated from the Mg {\tiny II} line, both for the average and night-wise estimations. The luminosities of the BLR, disc, and torus obtained from this line are $\log(L_{\mathrm{BLR}} \ \mathrm{[erg \ s^{-1}]}) = 44.97 \pm 0.10$, $\log(L_{\mathrm{disc}} \ \mathrm{[erg \ s^{-1}]}) = 45.97 \pm 0.10$, and $\log(L_{\mathrm{torus}} \ \mathrm{[erg \ s^{-1}]}) = 44.75 \pm 0.10$, respectively. We also estimated the size of the BLR and torus with Eqs.~(\ref{eq:blr_radius}) and (\ref{eq:torus_radius}), obtaining values of $r_{\mathrm{BLR}} = (3.06 \pm 0.33) \times 10^{17}$~cm and $r_{\mathrm{torus}} = (7.66 \pm 0.82) \times 10^{18}$~cm, again consistent with those derived from the Mg {\tiny II} line.

Finally, we were able to estimate the virial mass of the central black hole as a function of the FWHM and luminosity of the Mg~{\tiny II} line following \cite{shaw2012},
\begin{equation}
\log \left( \frac{M_{BH}}{M_\odot} \right) = a + b \log \left( \frac{L_{\mathrm{Mg {\tiny II}}}}{10^{44} \ \mathrm{erg \ s^{-1}}} \right) + 2 \log \left( \frac{FWHM_{\mathrm{Mg {\tiny II}}}}{\mathrm{km \ s^{-1}}} \right).
\label{eq:BH_mass}
\end{equation}
Here, $a$ and $b$ are empirically calibrated coefficients using reverberation mapping data \citep{shen2011}. When calculating the black hole mass from the emission line luminosity as in the present case, they adopt values of $a=1.70$ and $b=0.63$ for the Mg {\tiny II} line, as reported by \cite{shaw2012}. The average mass derived in this analysis is $\log(M_{BH}/M_{\odot})=8.36 \pm0.18$, consistent with the independent estimations reported by \cite{shaw2012}, \cite{ghisellini2015} and \cite{pandey2025}. 

\section{Characterization and variability of the non-thermal continuum}\label{sec5}

\begin{figure*}
    \centering
    \includegraphics[width=\linewidth]{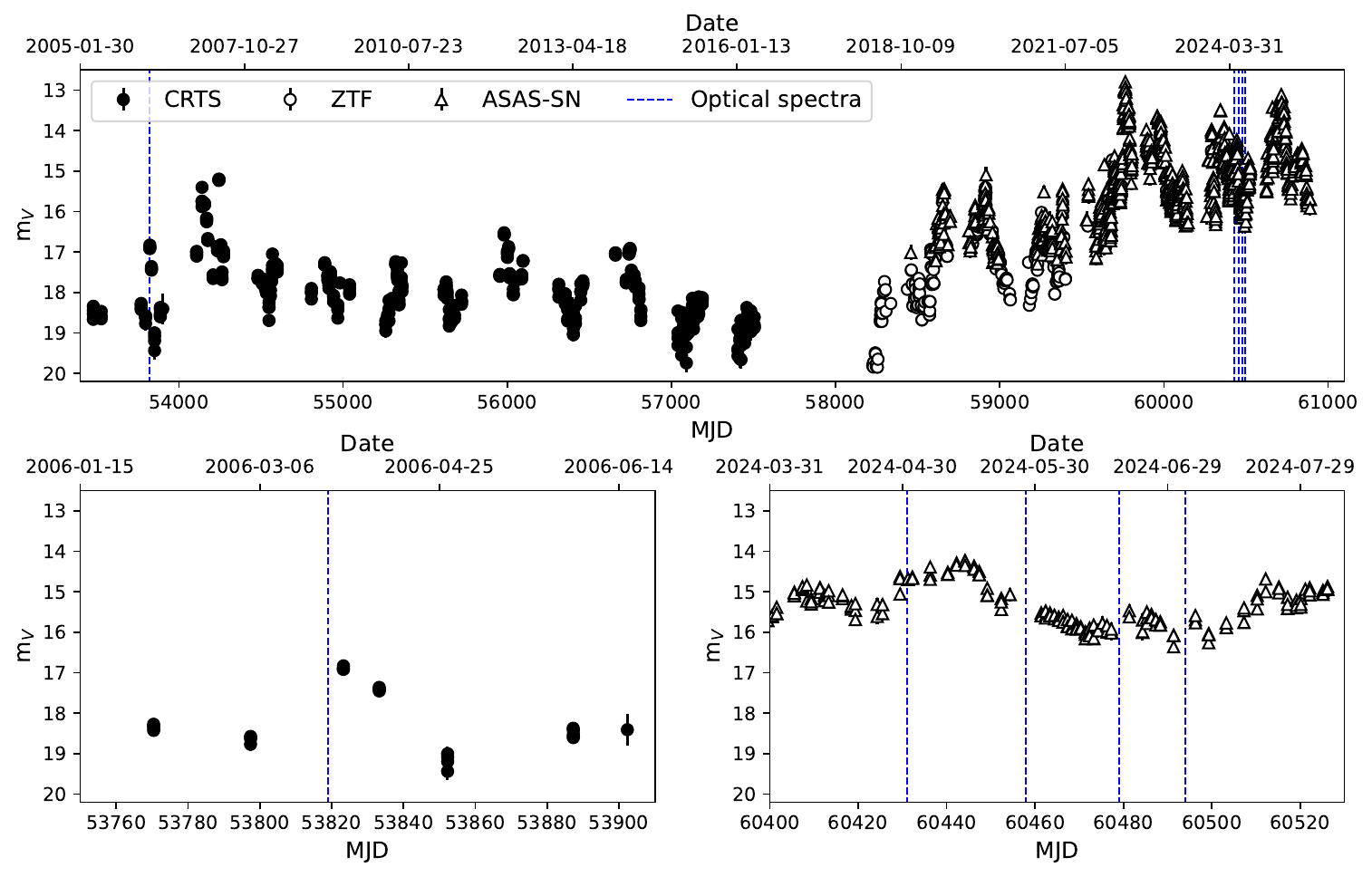}
    \caption{Historical optical $V$-band light curve of OP 313. Different markers represent data from the different databases used, as shown in the legend. Blue vertical lines highlight the dates of the spectroscopic observations. The top panel shows the complete light curve while the bottom left and right panels show the zoomed-in light curves around the dates of the SDSS and NOT/TNG spectra, respectively.}
    \label{fig:LCs}
\end{figure*}

The optical spectra were also used to characterize the variability of the continuum. From the flux changes between different observations, we measured a flux change of approximately a factor 10 from the faintest SDSS spectrum to the brightest one from NOT. In order to put into context the behaviour of OP~313 at the times of the spectroscopic observations, we have also compiled the historical optical $V$-band light curves. We retrieved data from the CATALINA Real-time Transient Survey \citep[CRTS,][]{drake2009}, that monitored the sky between 2005 to 2016 in search of transient events, and contains high-cadence data of several extragalactic objects, including blazars. We also included data in the $g$ band from the All-Sky Automated Survey for Supernovae \citep[ASAS-SN,][]{shappee2014,kochanek2017}, dedicated to supernova observations, but containing data of other astrophysical objects thanks to its all-sky scan capabilities. Additionally, we have downloaded data in the $g$ and $r$ bands from the Zwicky Transient Facility \citep[ZTF,][]{bellm2019,masci2019}, that monitors the northern sky compiling optical light curves of astrophysical objects since 2018. We have transformed the data in these two bands to the $V$ band to match that from the CRTS using the photometric transformations from Lupton (2005)\footnote{\url{https://classic.sdss.org/dr4/algorithms/sdssUBVRITransform.php\#Lupton2005}}. The $V$-band optical light curve is shown in Fig.~\ref{fig:LCs}.

The optical light curves of OP~313 between 2005 and 2025 show a historical variability amplitude of 7.06 magnitudes, from magnitude 19.85 at its faintest state to the maximum emission when the magnitude reached a value of 12.79. In particular, the closest observations to the spectrum taken by the SDSS show optical $V$-band magnitudes $\sim$17. On the other hand, during the spectroscopic observing campaign during 2024 the source shows brighter magnitudes ranging from $\sim$15 to $\sim$16. The older data taken by the CRTS clearly shows a more stable behaviour, with magnitude changes of $\sim$2 with the exception of a moderately bright flare reaching a magnitude 15.20. Starting on 2018, OP~313 started a phase of higher activity during which, apart from showing more and larger variations, a clear magnitude-increasing trend can be observed in the light curve.

We have also evaluated the spectral index change of the spectra from 2024 during the high emission state, comparing them with that obtained from the SDSS spectrum in 2006, when OP~313 was in a quiescent state. For this, we subtracted the estimated Mg {\tiny II} and C {\tiny III}] emission lines to the observed spectrum. Then, we fitted the remaining continuum emission with a power law shape defined as $F_{\lambda} \propto \lambda^{-\beta}$, where $\beta$ is the continuum spectral index in the wavelength space. We calculated as well the index in frequency space, that is, $F_{\nu} \propto \nu ^{-\alpha}$, as $\alpha = 2- \beta$. Finally, this was related with the spectral index of the underlying electron population responsible for this synchrotron emission as $p=2\alpha+1$ \citep{rybicki1986}. The spectral indices obtained for the different optical spectra are summarised in Table~\ref{tab:spectral_indices}.

\begin{table}
\centering
\caption{Spectral indices calculated from each optical spectrum in wavelength ($F_{\lambda}$, third column) and frequency ($F_{\nu}$, fourth column) representation, and the derived electron population spectral index (last column).}
\resizebox{\columnwidth}{!}{%
\label{tab:spectral_indices}
\begin{tabular}{ccccc}
\hline
Date & Instrument & $\beta$ & $\alpha$ & $p$ \\ \hline
2006-03-25 & SDSS        & $0.71 \pm 0.01$ & $1.29 \pm 0.01$ & $3.58 \pm 0.01$ \\ \hline
2024-05-01 & NOT/ALFOSC  & $0.48 \pm 0.01$ & $1.52 \pm 0.01$ & $4.04 \pm 0.01$ \\ \hline
2024-05-28 & TNG/DOLORES & $0.61 \pm 0.01$ & $1.39 \pm 0.01$ & $3.78 \pm 0.01$ \\ \hline
2024-05-28 & TNG/DOLORES & $0.66 \pm 0.01$ & $1.34 \pm 0.01$ & $3.68 \pm 0.01$ \\ \hline
2024-06-18 & NOT/ALFOSC  & $0.45 \pm 0.01$ & $1.55 \pm 0.01$ & $4.10 \pm 0.02$ \\ \hline
2024-07-03 & TNG/DOLORES & $0.43 \pm 0.01$ & $1.57 \pm 0.01$ & $4.14 \pm 0.01$ \\ \hline
2024-07-03 & TNG/DOLORES & $0.49 \pm 0.01$ & $1.51 \pm 0.01$ & $4.02 \pm 0.01$ \\ \hline
\end{tabular}
}
\end{table}

We observe that the spectral indices $\alpha$ of the synchrotron continuum are rather steep, with values ranging from $\sim$1.29 to $\sim$1.57 in $F_{\nu}$. This translates to also rather steep indices in the electron population of $p\sim3.6-4.1$, approximately. The hardest index, $p=3.58$, is observed in the archival SDSS spectrum, when the source was in low state and the continuum level was $\sim$10 times lower than that during the observations performed in 2024. Nevertheless, the variability of the indices is not dramatic over the different observations. The values are also consistent with those used for explaining the broadband emission of OP~313 in previous studies \citep[][CTAO-LST \& MAGIC Collaborations, submitted]{pandey2024,dar2025}. 

\begin{figure}
    \centering
    \includegraphics[width=\linewidth]{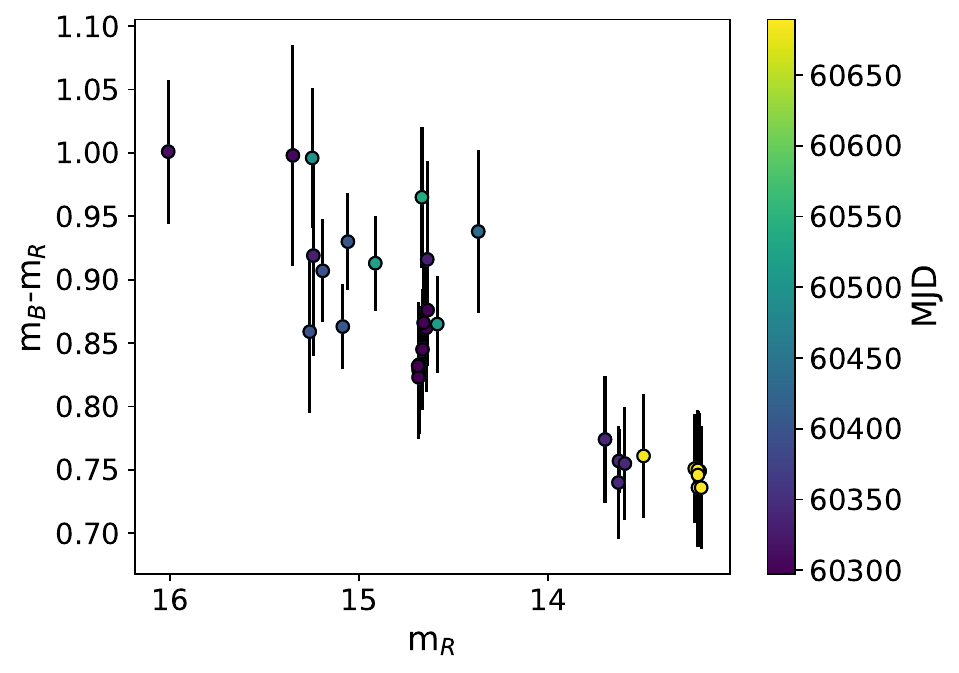}
    \includegraphics[width=\linewidth]{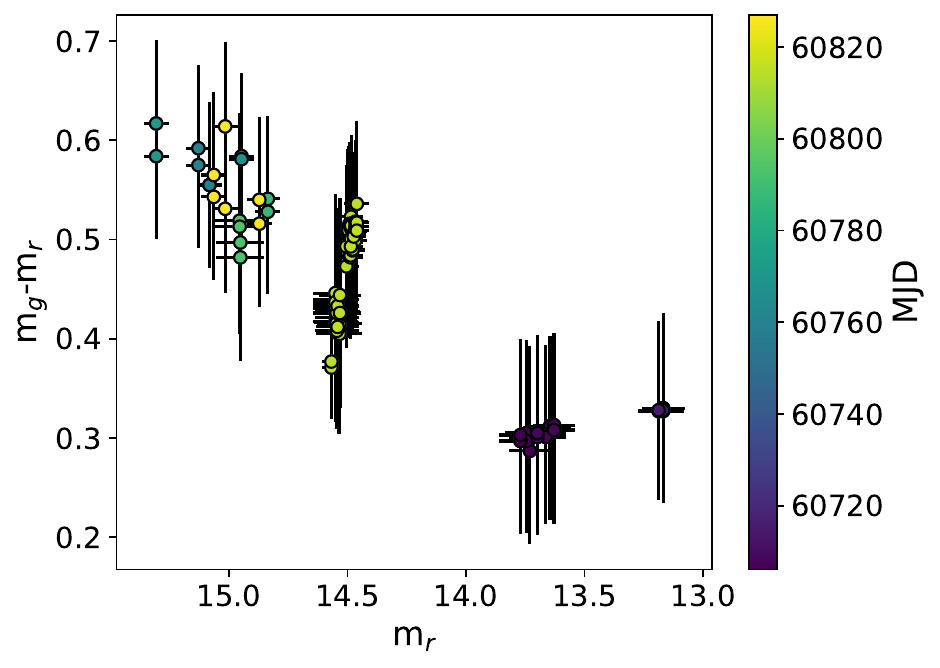}
    \caption{Colour variability analysis of OP~313 during the 2023-2025 period. \textit{Top:} $B-R$ colour with respect to the $R$-band magnitude. \textit{Bottom:} $g-r$ colour with respect to the $r$-band magnitude.}
    \label{fig:color_short}
\end{figure}

Moreover, we investigated possible trends in the spectral changes based on the colour variability derived from the multi-band optical light curves over several months. For this, we compiled observations coordinated with the IAC80 telescope at Teide Observatory (Tenerife, Spain), performed between December 2023 and May 2025, in the Johnson $BRI$ and Sloan $gri$ bands, analysed with standard photometry data reduction procedures. The light curves are presented in Appendix~\ref{sec:A3}. By evaluating the multi-band changes for the quasi-simultaneous observations, we observed a clear bluer-when-brighter (BWB) behaviour in weeks to months timescales, as represented in Fig.~\ref{fig:color_short}. In particular, we show here the $B-R$ and $g-r$ colour relations, however, we note the $R-I$, $I-B$, $r-i$, and $g-i$ colours also follow this BWB trend. These colour variations are rather mild, therefore in alignment with the relatively stable spectral indices derived from the optical spectra taken between May and July 2024. This BWB trend is contrary to the typical redder-when-brighter often associated with FSRQs \citep[see e.g.][]{zhang2015}, mostly due to the contributions of the thermal components to the optical emission. However, in many cases FSRQs have shown as well the BWB behaviour more commonly associated to BL Lac objects \citep{bonning2012,isler2017}, especially during high emission states such as the one shown here by OP~313, during which the non-thermal continuum outshines the thermal components and dominates the colour variability as it happens in BL Lacs \citep{otero2022}. Therefore, the colour variability observed for OP~313 is in line with the highly dominant synchrotron continuum between 2023 and 2025 during the very active period of this source.

\begin{figure}
    \centering
    \includegraphics[width=\linewidth]{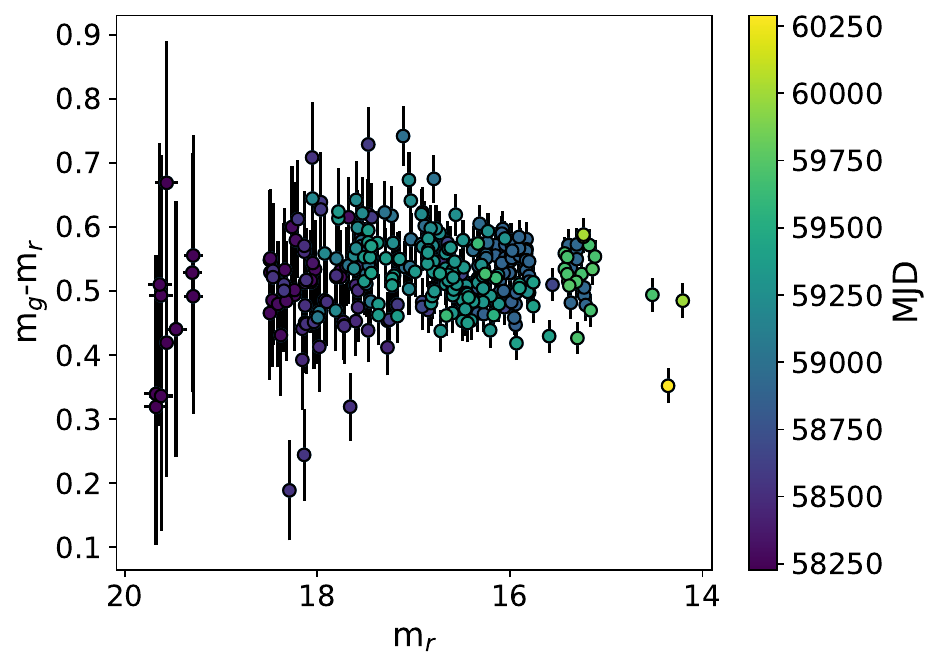}
    \caption{Long-term $g-r$ colour variability analysis of OP~313 with respect to the $r$-band magnitude from ZTF data between 2018 and 2024.}
    \label{fig:color_long}
\end{figure}

In addition, we compared the week-month timescale colour change with that over several years thanks to the $gr$ long-term light curves provided by the ZTF (also shown in Appendix~\ref{sec:A3}). The year timescale colour variability is represented in Fig.~\ref{fig:color_long}. Opposite to what we have observed on shorter timescales, here we see a rather stable colour of OP~313's optical emission. This behaviour has also been seen in the past for long-term variability of other blazars \citep[see for instance][]{raiteri2021,raiteri2021b,otero2024}. These achromaticity in the long-term variability of blazars has been ascribed to Doppler factor changes due to geometrical effects in the jet instead of intrinsic jet processes relative to the emitting region \citep[see][]{raiteri2021,raiteri2021b}.

\section{Discussion}\label{sec6}
The thermal emission from the accretion disc and dusty torus, as well as the BLR in blazars is typically assumed to remain constant over time, as potential changes in these components are typically expected on longer timescales than those accessible by regular observations \citep[see][and references therein]{otero2022}. Only in a handful of cases, observations have revealed evidences of significant variations, observable from the detection of changes in the intensity and profile of optical emission lines \citep[see][]{leon-tavares2013,chavushyan2020,hallum2022}. 

Owing to the results derived here, we observe that the behaviour of the Mg {\tiny II} is consistent with a constant flux over time. Changes between the different spectra can be ascribed to statistical and instrumental effects, as all results are consistent with the expected behaviour from a constant emission line. Therefore, the apparent change from one spectrum to the next is in reality due to an effect of blurring and dilution of the emission line by the highly variable synchrotron continuum, especially in those spectra from which the broad line component is completely covered by this continuum, coincident with the lowest --- but nevertheless still compatible --- measurements of the Mg {\tiny II} line flux.

This effect of dilution of the emission lines has been observed several times in blazars \citep[see e.g.][]{kang2025}. Given the bright BLR often observed in FSRQs, it is more commonly observed in BL Lacs that do not usually show emission lines except for their faintest states due to a fainter and smaller BLR \citep[][]{paiano2017}. However, several FSRQs have been observed to show this same behaviour in the past, where their optical lines are completely masked by a bright non-thermal continuum during flaring states \citep[see e.g.][]{mishra2021,becerra2021}. In some cases, this has been associated to a changing-look nature of the source, {that is}, a source that shows a dramatical change in its characteristics, for instance the BLR emission line profiles, accretion rate, or Compton dominance (ratio between the IC and synchrotron peaks, >1 for FSRQs and $\lesssim$1 for BL Lacs), behaving at times as a FSRQ and sometimes as a BL Lac \citep[see][and references therein]{ruan2014,kang2024}. These changes are often associated with clearly distinct emission states of the blazar. 

For the present case of OP~313 studied here, the characterization of the thermal components, the BLR, and the Mg {\tiny II} (see Tables~\ref{tab:emission_lines} and \ref{tab:thermal_components}) points towards a constant emission line profile diluted by the synchrotron emission rather than a BLR with changing properties. Moreover, the characterization of this emission line, the properties of these components, and the estimation of the black hole mass enabled the estimation of the Eddington luminosity and therefore, the ratio between the Eddington and the accretion disc luminosities, $\lambda = L_{AD}/L_{Edd}$. This ratio is often use a a measurement of the accretion efficiency, where values <$0.01$ are often associated with inefficient accretion regimes, expected from BL Lac objects, while larger values are typically found in FSRQs \citep{pandey2025}. Using the derived average black hole mass, we estimated a value for the Eddington luminosity of $L_{Edd}=2.82 \times 10^{46}$~erg~s$^{-1}$. This leads to values of $\lambda$ ranging from 0.14 and 0.47, and an average of $\lambda = 0.23 \pm 0.10$, consistent with an efficient accretion expected from a FSRQ such as OP~313. These characteristics, combined with a Compton dominance clearly >1 both in low and high emission states, as shown by previous studies \citep[see e.g.][CTAO-LST collaboration, submitted]{pandey2024, nievas-rosillo2025}, indicate that the source is not a changing-look blazar but a FSRQ whose BLR and optical emission lines get diluted and covered by a highly variable synchrotron continuum during flaring states.

Regarding the continuum emission, we measured a rather steep synchrotron spectrum, with values of the $F_{\nu}$ spectrum between $\alpha \sim1.3-1.6$ using a power law shape. These spectral indices are translated into an electron population spectral index ranging from $p\sim3.6$ to $p\sim$4.10. This shows that the low energy electrons dominate the particle population. These values could be indicating an electron population accelerated by shocks, that typically show values between $p=2$ and $p=3$, that is later radiatively cooled via synchrotron, introducing a steepening of the spectral steepening of $\sim$1 \citep{rybicki1986}. This causes that an injected particle population with index $p=2-3$ can appear as $p=3-4$, consistent with those measured here. 

\section{Conclusions}\label{sec7}
We have characterized the optical spectrum of the FSRQ OP~313 during a bright and active period between May and July 2024.
We compared its characteristics with those observed during the historically low emission state shown by this source in 2006 and represented by an SDSS spectrum.

\begin{itemize}
    \item We have calculated the average flux and luminosity of the Mg {\tiny II} line based on the detection of this spectral feature in six out of the seven spectra analysed here, with values of $F_{\mathrm{Mg \ {\tiny II}}} = (0.85 \pm 0.11)\times 10^{-14}$~erg~cm$^{-2}$~s$^{-1}$ and $\log(L_{\mathrm{Mg \ {\tiny II}}} \ \mathrm{[erg \ s^{-1}]}) = 43.72 \pm 0.07$, respectively. We have also detected the C {\tiny III}] emission line in five spectra, for which we measure and average flux and luminosity of $F_{\mathrm{C \ {\tiny III}]}} = (1.00 \pm 0.10)\times 10^{-14}$~erg~cm$^{-2}$~s$^{-1}$ and $\log(L_{\mathrm{C \ {\tiny III}]}} \ \mathrm{[erg \ s^{-1}]}) = 43.77 \pm 0.10$.

    \item We evaluated potential changes over time in the BLR through the characterization of the Mg {\tiny II} line. The results are compatible with a constant line flux and luminosity. Changes between different epochs in the |EW| $\sim 0.9$ to $\sim33$ can be ascribed to statistical fluctuations and effects of line blurring by the intense and highly variable continuum emission. This is also visible from the fluxes measured for the C {\tiny III}] line, consistent within errors.

    \item We constrained the size and luminosity of the BLR, accretion disc, and dusty torus through the Mg {\tiny II} line measurements. We obtained average luminosities $\log(L_{\mathrm{BLR}} \ \mathrm{[erg \ s^{-1}]}) = 44.91 \pm 0.19$, $\log(L_{\mathrm{disc}} \ \mathrm{[erg \ s^{-1}]}) = 45.91 \pm 0.19$, and $\log(L_{\mathrm{torus}} \ \mathrm{[erg \ s^{-1}]}) = 44.70 \pm 0.16$, and radii $r_{\mathrm{BLR}} = (2.76 \pm 0.53) \times 10^{17}$~cm and $r_{\mathrm{torus}} = (6.89 \pm 1.32) \times 10^{18}$~cm, consistent with previous estimations \citep{ghisellini2015, pandey2025}. The values obtained from the characterization of the C {\tiny III}] emission line are consistent within uncertainties.

    \item The black hole mass was also estimated. An average value of $\log(M_{BH}/M_{\odot})=8.36 \pm0.18$ was derived from the different spectra analysed here, consistent with the estimations from \cite{shaw2012}, \cite{ghisellini2015} and \cite{pandey2025}.

    \item The colour variability shows a BWB behaviour on week to month timescales due to the dominance of the synchrotron emission during the period evaluated. On the other hand, the long-term colour variations show a quasi-achromatic evolution.

    \item The particle population shows steep spectral indices with values between $\sim$3.6 and $\sim$4.10, that could indicate a shock accelerated and radiatively cooled electron population \citep{rybicki1986}.

    \item We estimated an Eddington luminosity of $L_{Edd}=2.82 \times 10^{46}$~erg~s$^{-1}$. The average ratio between the accretion disc and Eddington luminosities is $\lambda = 0.23 \pm 0.10$, indicative of an efficient accretion rate.
    
\end{itemize}

Overall, the characteristics discussed here --- constant emission line, $\lambda = 0.23 \pm 0.10$ and Compton dominance >1 --- disfavour a changing-look nature of OP~313, are in agreement with a FSRQ-like nature of this source, with a strong and variable non-thermal continuum masks the optical emission lines. This is consistent with the findings reported by \cite{pandey2025}.

\begin{acknowledgements}
{We sincerely thank the anonymous referee for their revision of the manuscript and the helpful and positive feedback.}

J.O.-S. acknowledges founding from the Istituto Nazionale di Fisica Nucleare (INFN) Cap. U.1.01.01.01.009.

M.N.R. and J.A.-P. acknowledge support from the Agencia Estatal de Investigación del Ministerio de Ciencia, Innovación y Universidades (MCIU/AEI) under grant PARTICIPACIÓN DEL IAC EN EL EXPERIMENTO AMS and the European Regional Development Fund (ERDF) with reference PID2022-137810NB-C22/DIO 10.13039/501100011033.

M.N.R. acknowledges funding from the Viera y Clavijo Programme of the University of La Laguna and the Government of the Canary Islands.

J.A.-P. acknowledges financial support from the Spanish Ministry of Science and Innovation (MICINN) through the Spanish State Research Agency, under Severo Ochoa Centres of Excellence Programme 2020-2024 (CEX2019-000920-S).

This article is based on observations made with the IAC80 operated on the island of Tenerife by the Instituto de Astrofísica de Canarias in the Spanish Observatorio del Teide.

This article is based on observations made in the Observatorios de Canarias del IAC with the NOT operated on the island of La Palma in the Observatorio del Roque de Los Muchachos. The observations were performed through the Service Night time, with Service Proposals SST2024-666 and SST2024-660. The data presented here were obtained [in part] with ALFOSC, which is provided by the Instituto de Astrofísica de Andalucía (IAA) under a joint agreement with the University of Copenhagen and NOT.

Based on observations made with the Italian Telescopio Nazionale Galileo (TNG) operated on the island of La Palma by the Fundación Galileo Galilei of the INAF (Istituto Nazionale di Astrofisica) at the Spanish Observatorio del Roque de los Muchachos of the Instituto de Astrofisica de Canarias. The observations were performed under Service Night proposal SST2024-665.

Funding for the Sloan Digital Sky Survey V has been provided by the Alfred P. Sloan Foundation, the Heising-Simons Foundation, the National Science Foundation, and the Participating Institutions. SDSS acknowledges support and resources from the Center for High-Performance Computing at the University of Utah. SDSS telescopes are located at Apache Point Observatory, funded by the Astrophysical Research Consortium and operated by New Mexico State University, and at Las Campanas Observatory, operated by the Carnegie Institution for Science. The SDSS web site is \url{www.sdss.org}.

SDSS is managed by the Astrophysical Research Consortium for the Participating Institutions of the SDSS Collaboration, including Caltech, The Carnegie Institution for Science, Chilean National Time Allocation Committee (CNTAC) ratified researchers, The Flatiron Institute, the Gotham Participation Group, Harvard University, Heidelberg University, The Johns Hopkins University, L'Ecole polytechnique f\'{e}d\'{e}rale de Lausanne (EPFL), Leibniz-Institut f\"{u}r Astrophysik Potsdam (AIP), Max-Planck-Institut f\"{u}r Astronomie (MPIA Heidelberg), Max-Planck-Institut f\"{u}r Extraterrestrische Physik (MPE), Nanjing University, National Astronomical Observatories of China (NAOC), New Mexico State University, The Ohio State University, Pennsylvania State University, Smithsonian Astrophysical Observatory, Space Telescope Science Institute (STScI), the Stellar Astrophysics Participation Group, Universidad Nacional Aut\'{o}noma de M\'{e}xico, University of Arizona, University of Colorado Boulder, University of Illinois at Urbana-Champaign, University of Toronto, University of Utah, University of Virginia, Yale University, and Yunnan University.

The CSS survey is funded by the National Aeronautics and Space Administration under Grant No. NNG05GF22G issued through the Science Mission Directorate Near-Earth Objects Observations Program.  The CRTS survey is supported by the U.S.~National Science Foundation under grants AST-0909182.

Based on observations obtained with the Samuel Oschin Telescope 48-inch and the 60-inch Telescope at the Palomar Observatory as part of the Zwicky Transient Facility project. ZTF is supported by the National Science Foundation under Grants No. AST-1440341 and AST-2034437 and a collaboration including current partners Caltech, IPAC, the Oskar Klein Center at Stockholm University, the University of Maryland, University of California, Berkeley, the University of Wisconsin at Milwaukee, University of Warwick, Ruhr University, Cornell University, Northwestern University and Drexel University. Operations are conducted by COO, IPAC, and UW.

This work makes use of data from the All-Sky Automated Survey for Supernovae (ASAS-SN).

\end{acknowledgements}

\bibliographystyle{aa}  
\bibliography{aa}

\begin{appendix} 

\onecolumn

\section{Minimum measurable EW}\label{sec:A1}
Following the procedure from \cite{becerra2021}, we calculated the minimum detectable EW for which we can claim that an emission line is significantly resolved. This $\mathrm{EW_{min}}$ can be expressed as
\begin{equation}
\mathrm{EW_{min}}=n_{\sigma} \sigma(\text{EW}) \simeq n_{\sigma} \frac{\sqrt{(N_{pix})}RMS_c \delta\lambda}{F_c(\lambda)} \simeq n_{\sigma} \frac{\sqrt{\Delta \lambda \delta \lambda}}{S/N_c},
\label{eq:EW_min}
\end{equation}
where $N_{pix}$ is the number of pixels containing the spectral feature, $RMS_c$ is the continuum noise, $\delta\lambda$ is the spectral dispersion, $F_c(\lambda)$ corresponds to the continuum flux, $\Delta \lambda$ is the wavelength window, and $S/N_c$ is the S/N of the continuum.

We fitted the continuum adjacent to an emission line by a {linear function}, that is then subtracted to the spectrum. The standard deviation of the residuals was then used for estimating the continuum noise $RMS_c$.
Assuming that fluctuations from neighbour pixels are uncorrelated, we could estimate the noise per pixel as the fluctuations of continuum adjacent pixels. The continuum was interpolated in the spectral range of the line, and a set of MC simulations were performed using as noise a random normal distribution with the noise per pixel estimated before. Then, we measured the EW error as the width of the EW distribution obtained from the MC simulations. This EW uncertainty was then used to calculate the $\mathrm{EW_{min}}$ as the 3$\sigma$ confidence level ($n_{\sigma}=3$). In Appendices~\ref{sec:A1.1} (Fig.~\ref{fig:MgII_EWmin}) and ~\ref{sec:A1.2} (Fig.~\ref{fig:CIII_EWmin}) we show the $\mathrm{EW_{min}}$ for the Mg {\tiny II} and C {\tiny III}] lines.

\subsection{Mg II emission line}\label{sec:A1.1}

\begin{figure*}[h]
\centering
\includegraphics[width=0.315\textwidth]{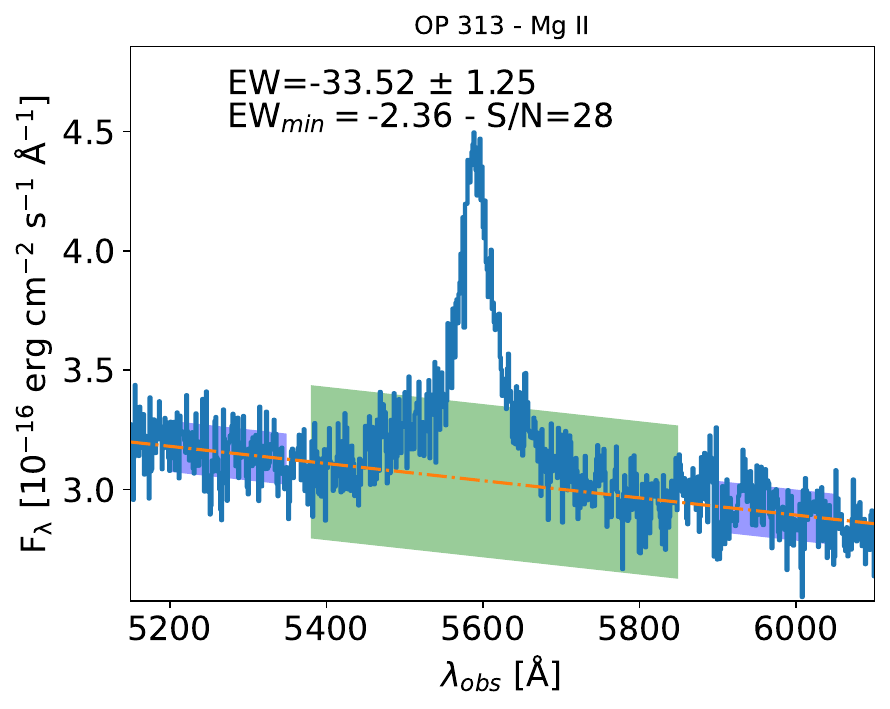}
\includegraphics[width=0.315\textwidth]{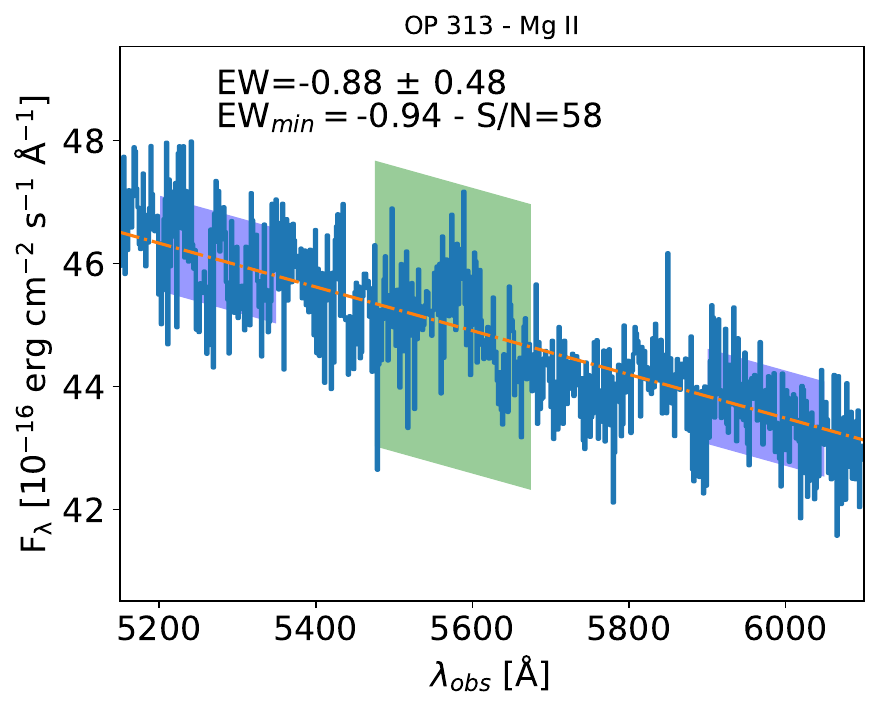}
\includegraphics[width=0.315\textwidth]{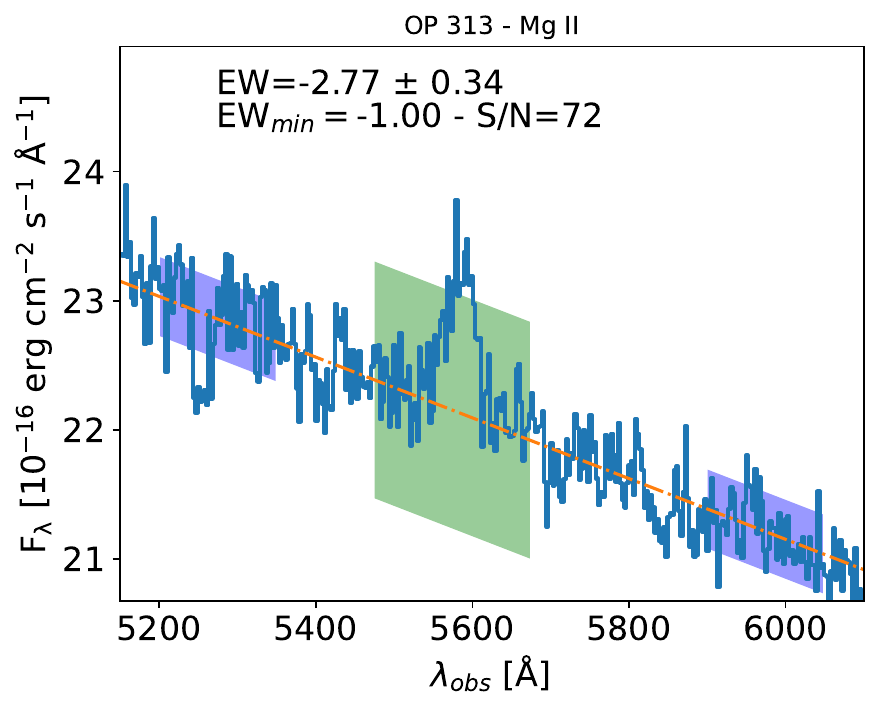}
\includegraphics[width=0.315\textwidth]{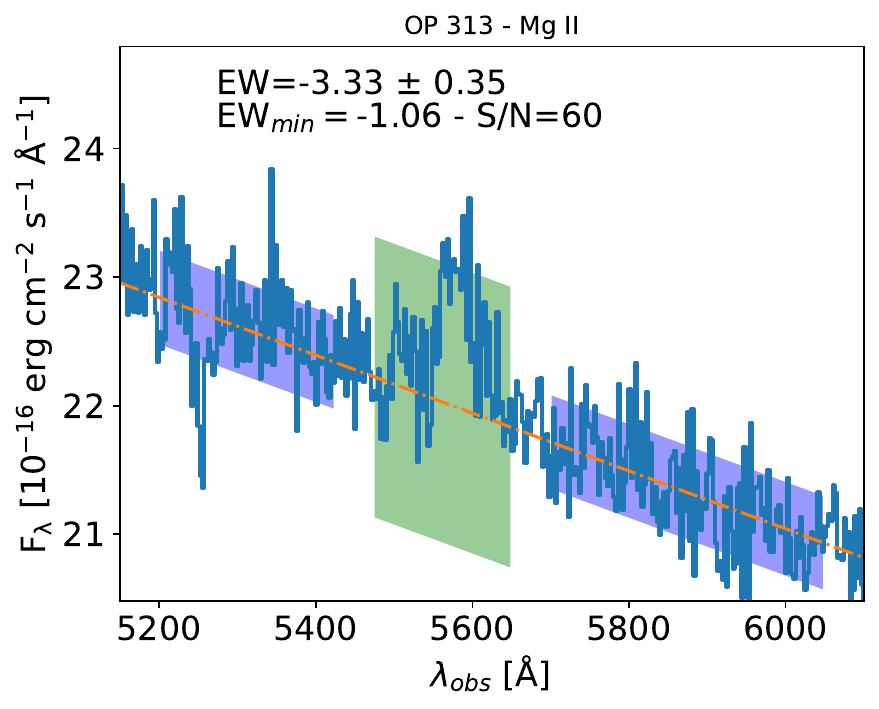}
\includegraphics[width=0.315\textwidth]{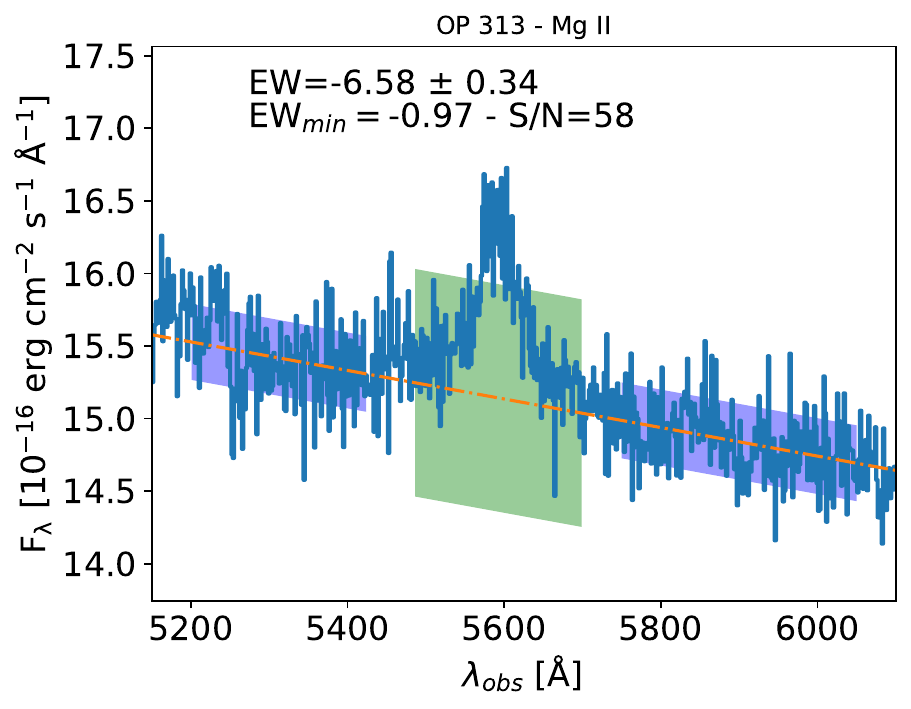}
\includegraphics[width=0.315\textwidth]{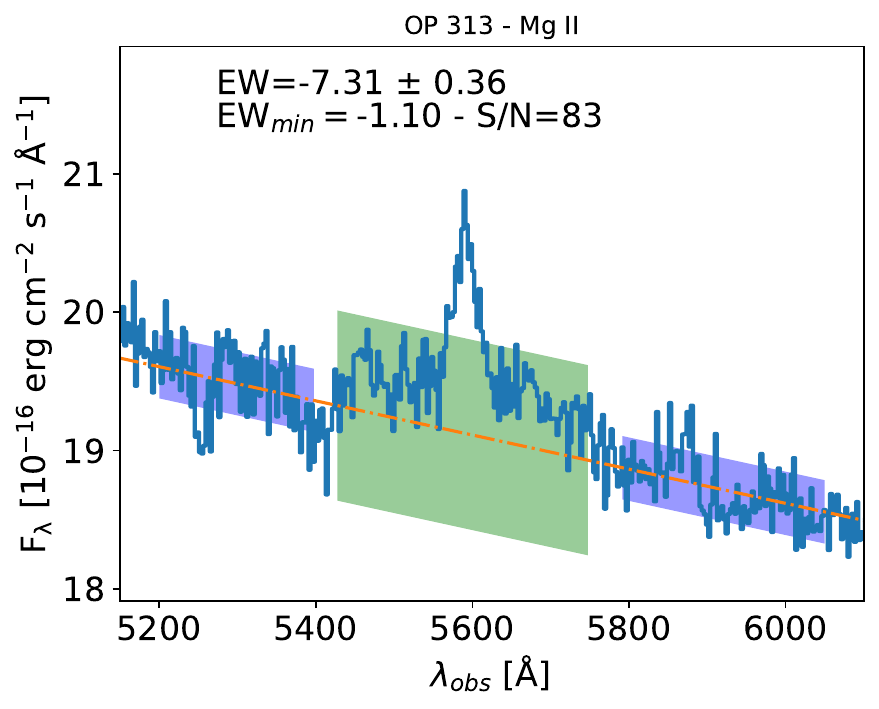}
\includegraphics[width=0.315\textwidth]{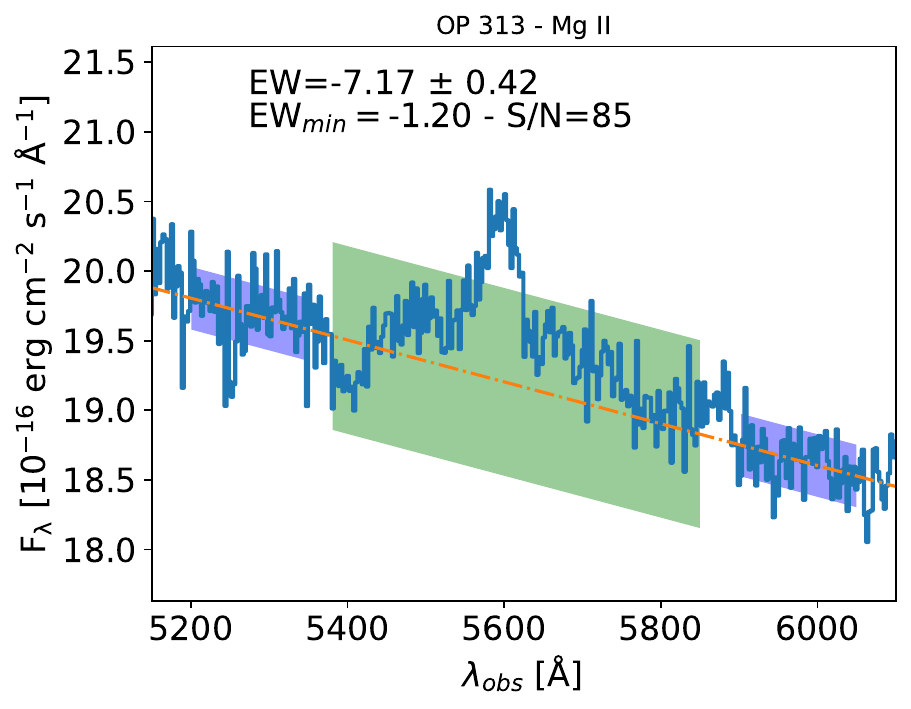}
\caption{Measurement of the $\mathrm{EW_{min}}$ for the Mg {\tiny II} emission line of the OP~313 spectra analysed in this work. \textit{Top left}: SDSS spectrum from 2006. \textit{Top middle}: NOT spectrum from May 1. \textit{Top right}: TNG spectrum from May 28 (first exposure). \textit{Centre left}: TNG spectrum from May 28 (second exposure). \textit{Centre middle}: NOT spectrum from June 18. \textit{Centre right}: TNG spectrum from July 3 (first exposure). \textit{Bottom}: TNG spectrum from July 3 (second exposure). Continuum regions are represented in blue and their vertical size corresponds to the 1$\sigma$ noise value. The window of the emission line is highlighted in green and its size represents a 3$\sigma$ noise value. The continuum interpolation is represented by the dotted-dashed line.}
\label{fig:MgII_EWmin}
\end{figure*}

\subsection{C III] emission line}\label{sec:A1.2}

\begin{figure*}[h]
\centering
\includegraphics[width=0.315\textwidth]{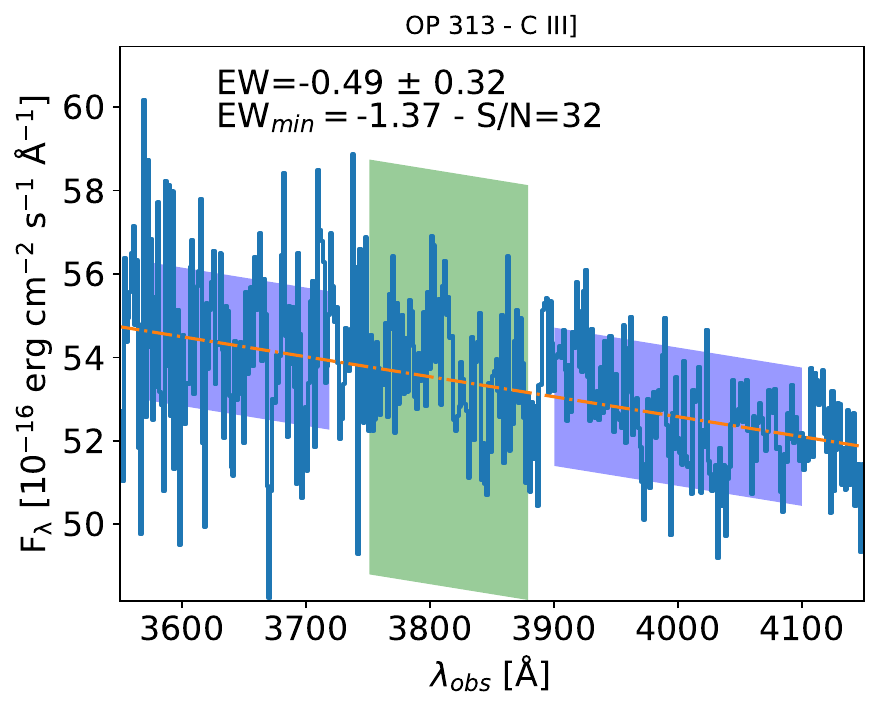}
\includegraphics[width=0.315\textwidth]{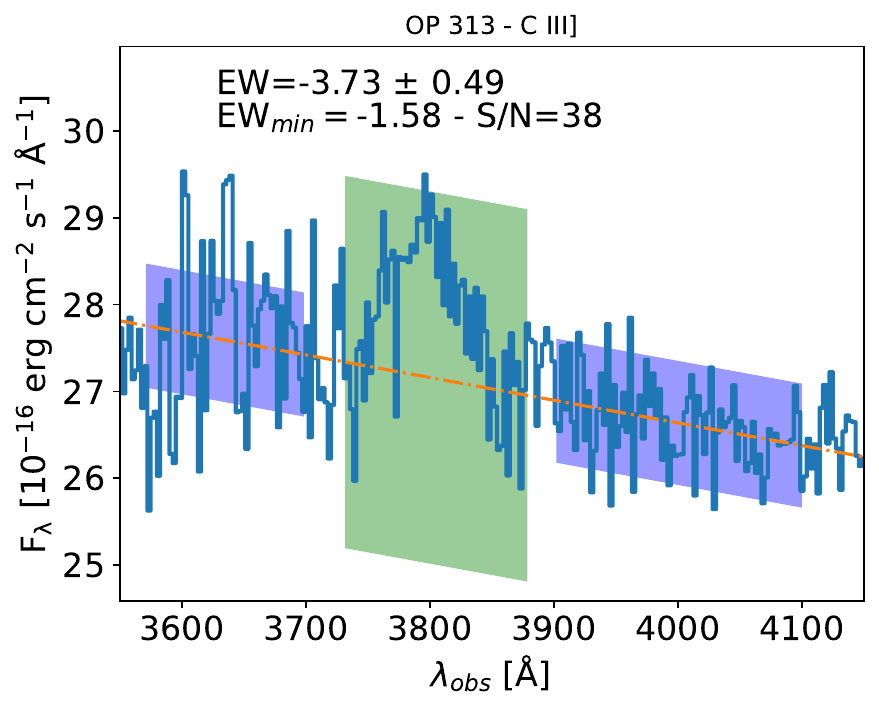}
\includegraphics[width=0.315\textwidth]{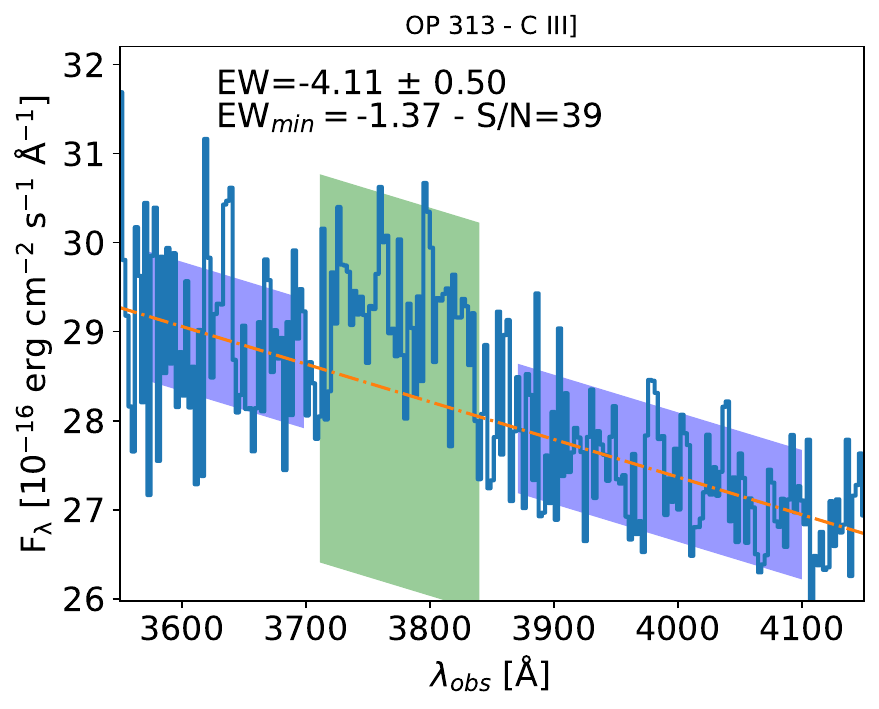}
\includegraphics[width=0.315\textwidth]{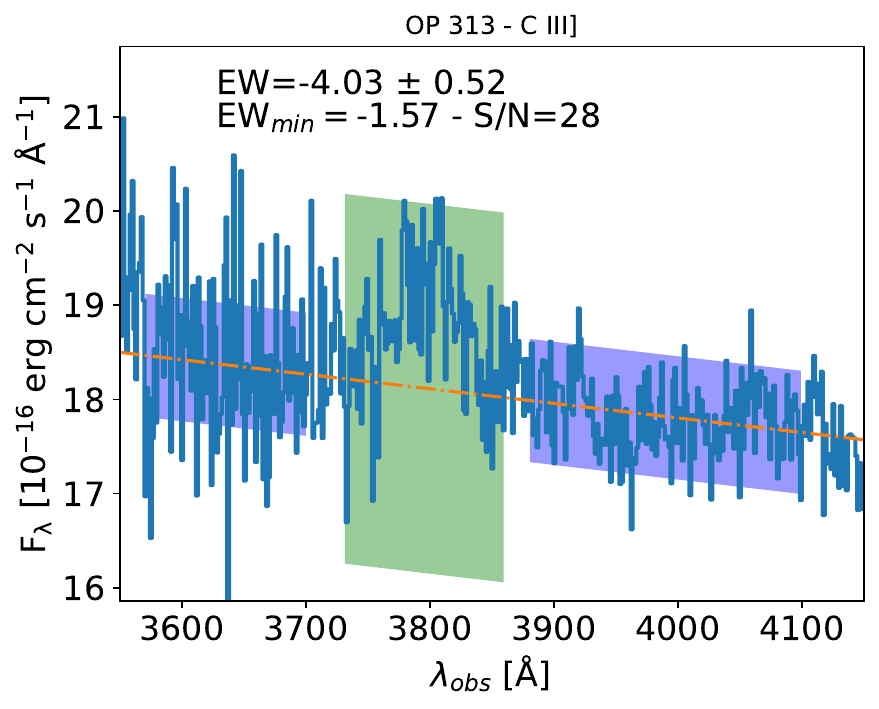}
\includegraphics[width=0.315\textwidth]{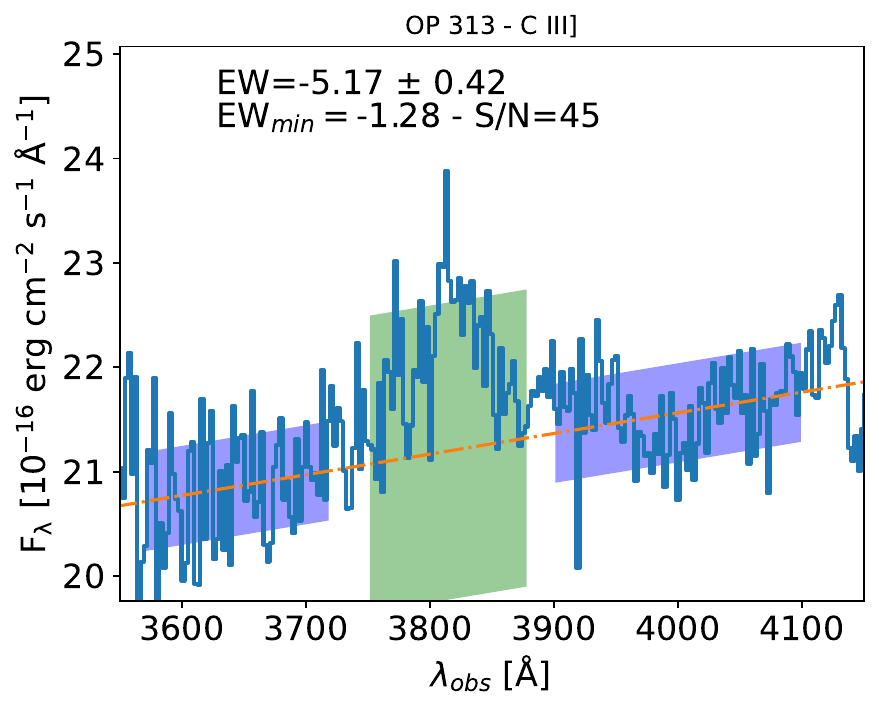}
\includegraphics[width=0.315\textwidth]{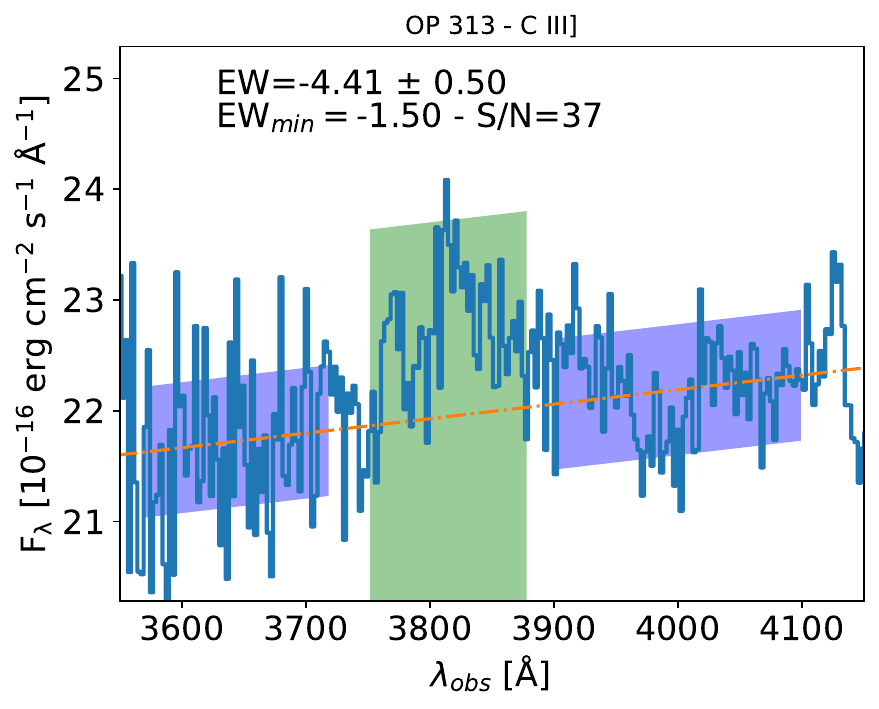}
\caption{Measurement of the $\mathrm{EW_{min}}$ for the C {\tiny III}] emission line of the OP~313 spectra analysed in this work. \textit{Top left}: NOT spectrum from May 1. \textit{Top middle}: TNG spectrum from May 28 (first exposure). \textit{Top right}: TNG spectrum from May 28 (second exposure). \textit{Bottom left}: NOT spectrum from June 18. \textit{Bottom middle}: TNG spectrum from July 3 (first exposure). \textit{Bottom right}: TNG spectrum from July 3 (second exposure).  Continuum regions are represented in blue and their vertical size corresponds to the 1$\sigma$ noise value. The window of the emission line is highlighted in green and its size represents a 3$\sigma$ noise value. The continuum interpolation is represented by the dotted-dashed line.}
\label{fig:CIII_EWmin}
\end{figure*}

\newpage

\section{Emission line profile fitting}\label{sec:A2}

\subsection{Mg II emission line}

Here we present the fits of the Mg {\tiny II} emission line for the different spectra taken during 2024, and introduced in Sects.~\ref{sec2}~and~\ref{sec3}. We include the Gaussian fit of the spectrum from May 1 despite not being significantly detected, since its EW was calculated and used for the evaluation of the expected EW from a constant emission line, presented in Fig.~\ref{fig:EW_continuum}. The spectra taken on May 1, May 28, and June 18 were modelled with a single Gaussian plus a linear function to model the continuum emission. On the other hand, the spectra from the night of June 3 were modelled with two Gaussians to account for the resolved narrow and broad emission line components, plus the linear function.

\begin{figure*}[h]
\centering
\includegraphics[width=0.33\textwidth]{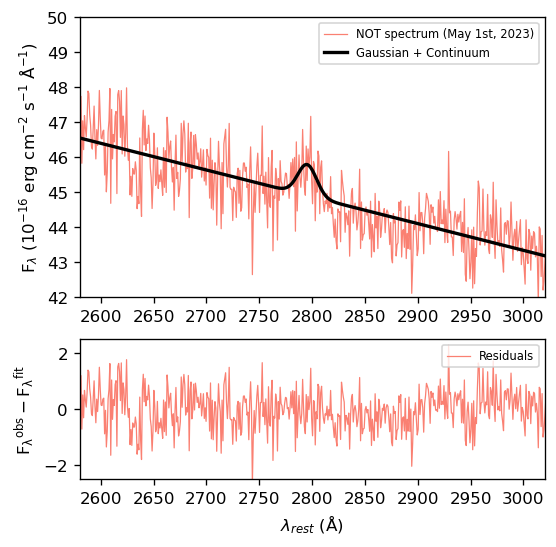}
\includegraphics[width=0.33\textwidth]{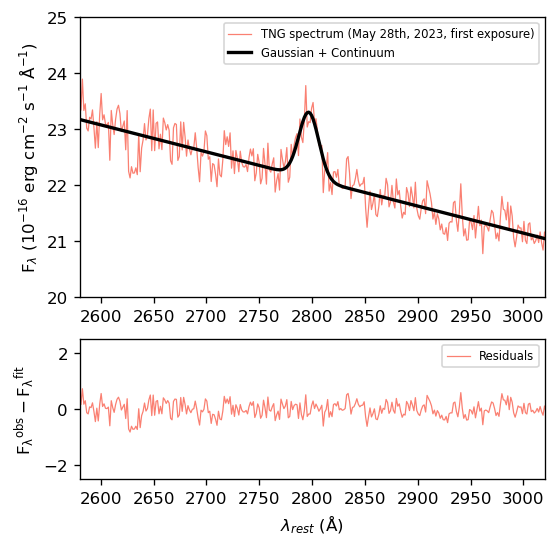}
\includegraphics[width=0.33\textwidth]{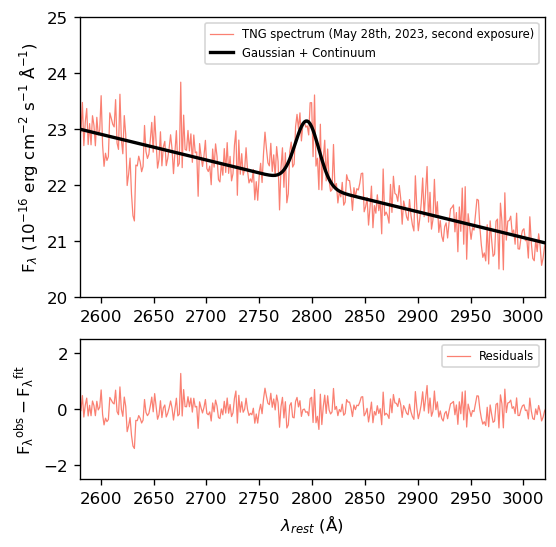}
\includegraphics[width=0.33\textwidth]{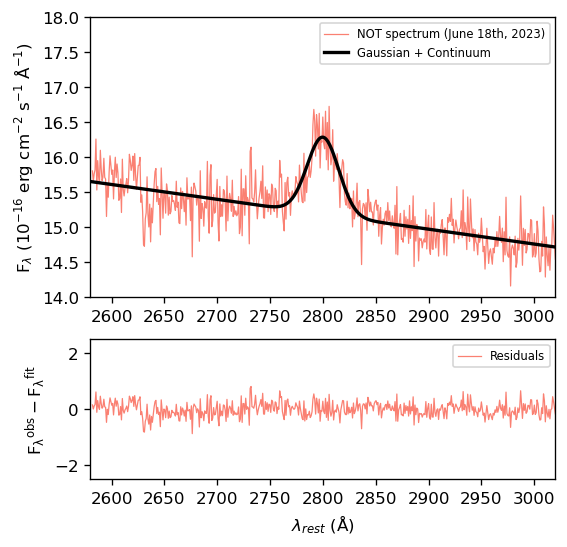}
\includegraphics[width=0.33\textwidth]{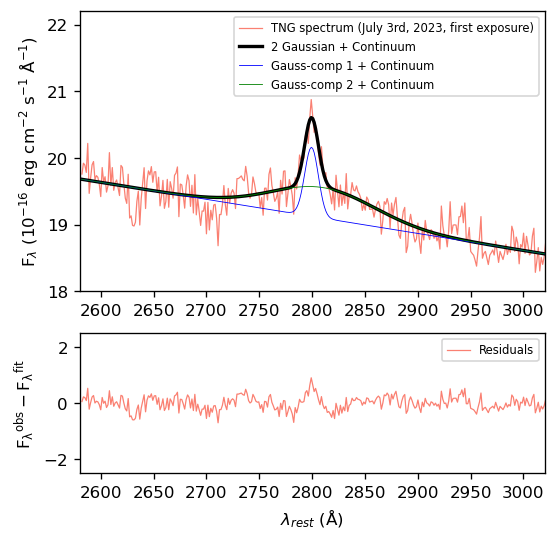}
\includegraphics[width=0.33\textwidth]{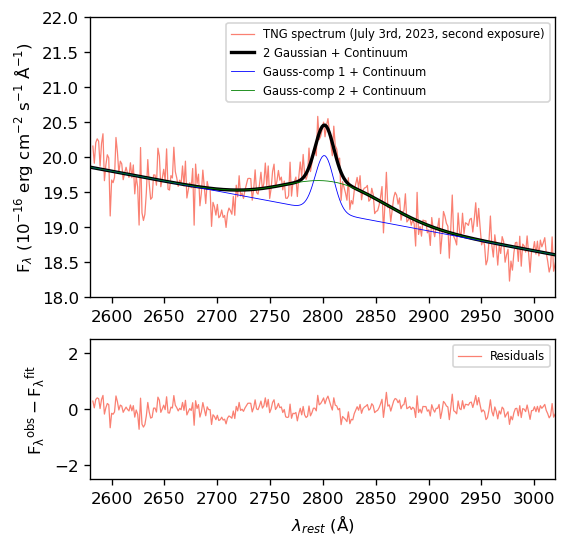}
\caption{Modelling of the Mg {\tiny II} line profile for the spectra taken by the NOT and TNG telescopes. \textit{Top left}: NOT spectrum from May 1. \textit{Top middle}: TNG spectrum from May 28 (first exposure). \textit{Top right}: TNG spectrum from May 28 (second exposure). \textit{Bottom left}: NOT spectrum from June 18. \textit{Bottom middle}: TNG spectrum from July 3 (first exposure). \textit{Bottom right}: TNG spectrum from July 3 (second exposure). For the spectra fitted with two Gaussian components plus a linear function, we represent the sum of the narrow component and the continuum in blue, the broad component and the continuum in red, and the total fitted model in black. For the spectra fitted with one Gaussian component plus a linear function we represent the total fitted model in black. The bottom panels show the residuals of the fits.}
\label{fig:MgII_line_all}
\end{figure*}

\newpage

\subsection{C III] emission line}
In this appendix we include the fits of the C {\tiny III}] emission line for those spectra in which the line was significantly resolved from the continuum emission (see Appendix~\ref{sec:A1.2} and Fig.~\ref{fig:CIII_EWmin}). The spectra were modelled with a single Gaussian function plus a linear function to model the continuum emission.

\begin{figure*}[h]
\centering
\includegraphics[width=0.33\textwidth]{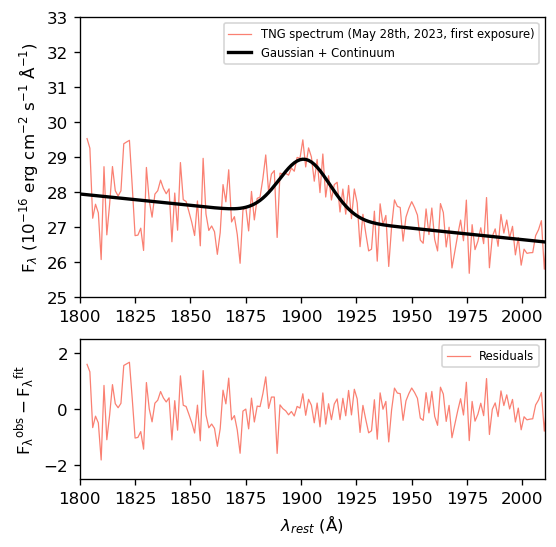}
\includegraphics[width=0.33\textwidth]{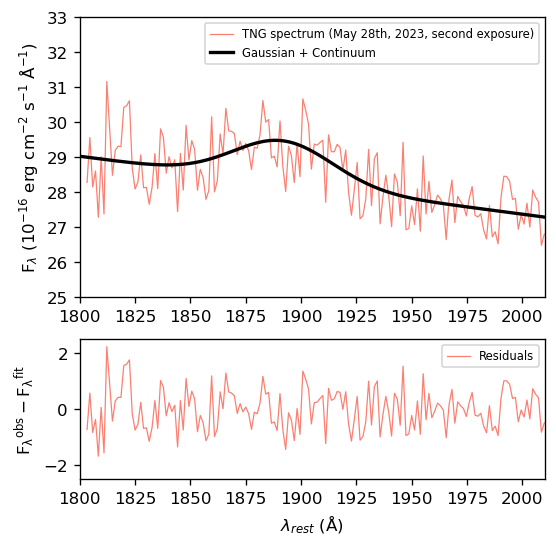}
\includegraphics[width=0.33\textwidth]{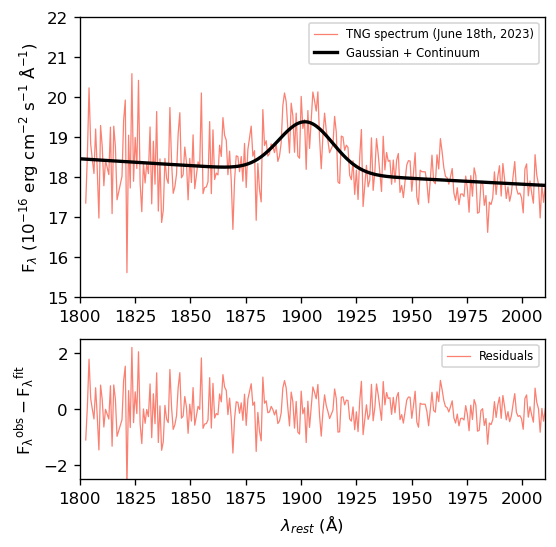}
\includegraphics[width=0.33\textwidth]{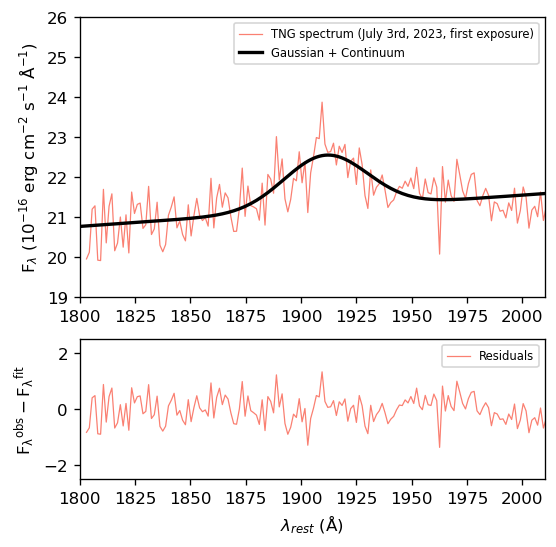}
\includegraphics[width=0.33\textwidth]{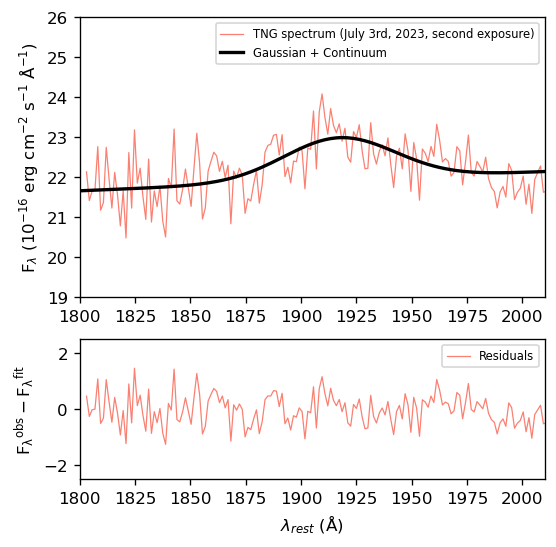}
\caption{Modelling of the C {\tiny III}] line profile for the spectra taken by the NOT and TNG telescopes. \textit{Top left}: TNG spectrum from May 28 (first exposure). \textit{Top middle}: TNG spectrum from May 28 (second exposure). \textit{Top right}: NOT spectrum from June 18. \textit{Bottom left}: TNG spectrum from July 3 (first exposure). \textit{Bottom right}: TNG spectrum from July 3 (second exposure). The spectra are fitted with one Gaussian component plus a linear function. The total fitted model is represented in black. The bottom panels show the residuals of the fits.}
\label{fig:CIII_line_all}
\end{figure*}

\newpage

\section{Multi-band optical light curves}\label{sec:A3}

\begin{figure*}[h]
\centering
\includegraphics[width=\textwidth]{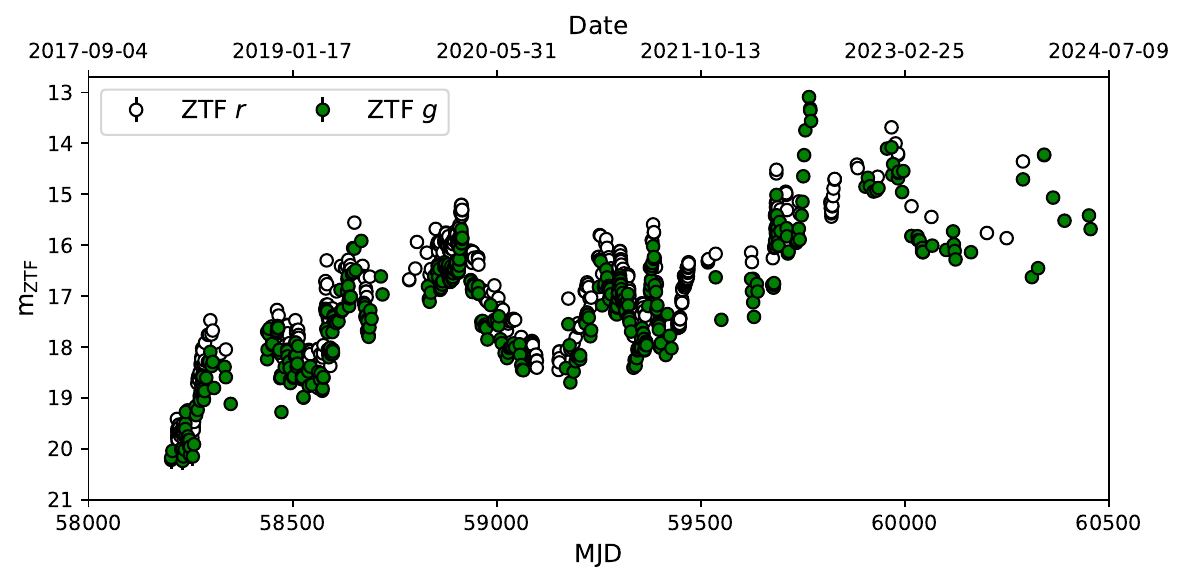}
\caption{ZTF long-term light curves of OP~313 from 2018 to 2024 in the Sloan $g$ (green markers) and $r$ (white markers) bands.}
\label{fig:ZTF_LC}
\end{figure*}

\begin{figure*}[h]
\centering
\includegraphics[width=\textwidth]{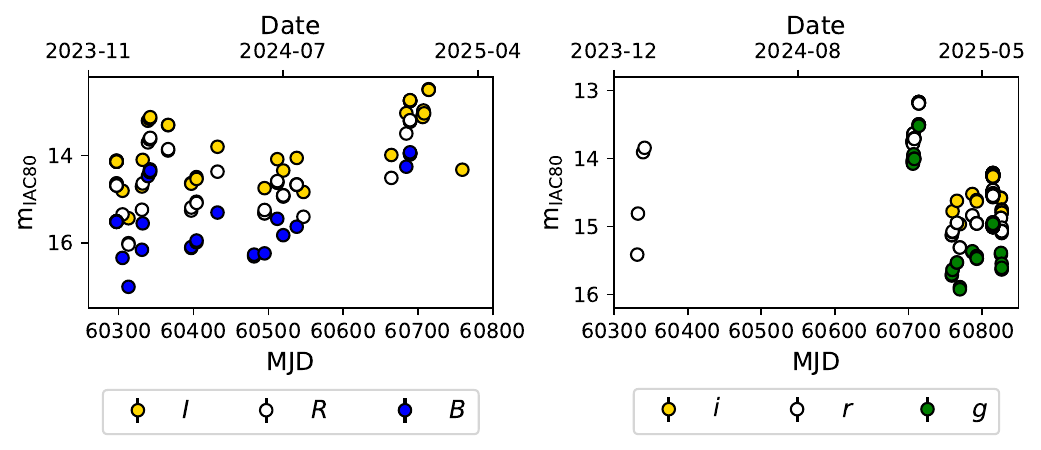}
\caption{IAC80 multi-band optical light curves of OP~313 in different optical bands from December 2023 to May 2025. \textit{Left:} Johnson $B-$band (blue markers), $R-$band (white markers), and $I-$band (gold markers) optical light curves. \textit{Right:} Sloan $g-$band (green markers), $r-$band (white markers), and $i-$band (gold markers) optical light curves. }
\label{fig:IAC80_LC}
\end{figure*}

In this appendix we show the long term multi-band light curves of OP~313 from ZTF in the Sloan $g$ and $r$ bands (Fig.~\ref{fig:ZTF_LC}) from 2018 to 2024; and the month-scale multi-band light curves resulting from the data of the IAC80 telescope in the Johnson $B$, $R$, and $I$ bands (Fig.~\ref{fig:IAC80_LC}, left panel), and the Sloan $g$, $r$, and $i$ bands (Fig.~\ref{fig:IAC80_LC}, right panel).

\end{appendix}

\end{document}